\documentclass[12pt,sort&compress]{elsarticle}

\usepackage[utf8]{inputenc}

\usepackage{bm}
\usepackage{mathrsfs}
\usepackage{amsmath}
\usepackage{amsfonts}
\usepackage{amssymb}
\usepackage{amsthm}
\usepackage{upgreek}
\usepackage{txfonts}
\usepackage{sansmath}
\usepackage{accents}

\usepackage{graphicx}
\usepackage{array}
\usepackage{booktabs}
\usepackage{multirow}
\usepackage{arydshln}
\usepackage{siunitx}

\usepackage{float}          
\usepackage{caption}
\usepackage{subcaption}
\usepackage{placeins}

\usepackage[dvipsnames]{xcolor}

\definecolor{shadecolor}{cmyk}{0,0,0,0.03}

\usepackage[
    top=1in,
    bottom=1in,
    left=1in,
    right=1in
]{geometry}

\usepackage{setspace}
\usepackage{fancyhdr}
\usepackage{multicol}
\usepackage{enumitem}
\usepackage{lipsum}
\usepackage{nth}

\allowdisplaybreaks

\usepackage{url}
\usepackage[
    pdfborder={0 0 0},
    colorlinks=true,
    allcolors=blue
]{hyperref}

\theoremstyle{definition}

\usepackage{prettyref}

\newrefformat{fig}{Figure~\ref{#1}}
\newrefformat{Fig}{Fig.~\ref{#1}}

\newrefformat{eq}{Eq.~(\ref{#1})}
\newrefformat{Eq}{Eq.~(\ref{#1})}

\newrefformat{sec}{Section~\ref{#1}}

\newrefformat{tab}{Table~\ref{#1}}
\newrefformat{Tab}{Table~\ref{#1}}

\newrefformat{alg}{Algorithm~\ref{#1}}
\newrefformat{Alg}{Algorithm~\ref{#1}}

\usepackage{algorithm}
\usepackage{algpseudocode}

\usepackage{lineno}

\journal{Journal of Fluid Mechanics}

\begin{document}

\begin{frontmatter}







\title{{\normalsize\textit{This manuscript has been submitted to the Journal of Fluid Mechanics.}
\url{https://www.cambridge.org/core/journals/journal-of-fluid-mechanics}}\\[0.5em]
Computational and experimental study of multi-grain-size sediment transport and sorting process}


\author[sbu]{Mehrshad Gholami Anjiraki}
\author[sbu]{Gabriel Narh}
\author[sbu]{Jonathan Craig}
\author[sbu]{Mustafa Meriç Aksen}
\author[sbu]{Samin Shapourmiandouab}
\author[sbu]{Ali Farhadzadeh}
\author[sbu]{Tao Xiang}

\author[sbu]{Ali Khosronejad\corref{cor1}}
\author[sbu]{Ali Khosronejad\corref{cor1}}
\ead{ali.khosronejad@stonybrook.edu}
\cortext[cor1]{Corresponding author}

\address[sbu]{Department of Civil Engineering, Stony Brook University, Stony Brook, NY 11794, USA}

\begin{abstract}
\noindent Successful prediction of the interaction between turbulent flows and the ecosystem of rivers is essential for flood management, development of mitigation strategies, and infrastructure protection. Natural riverbeds consist of bed materials of multiple grain sizes that interact with turbulence differently, leading to spatial classification of particle sizes. Such particle sorting processes can only be captured by models capable of handling multiple grain sizes. We present a numerical framework that couples large-eddy simulation with a multi-grain-size sediment transport module to track multiple grain size classes, their local fractions within the bed-load layer, and bed evolution. A laboratory experiment was conducted in a flume with multi-size bed material to examine sediment particle redistribution and sorting. The measured data for sediment grain size distributions and bed elevations were utilized to validate the developed numerical framework. The results of the multi-grain-size sediment transport model for bed deformation were then compared against those of a uniform sediment transport model, demonstrating that the former reduces the prediction errors relative to the experimental measurements from $37.98\%$ to $3.83\%$ for maximum scour depth and from $48.35\%$ to $11.54\%$ for maximum deposition height. The validated numerical framework was also applied to a number of large-scale meandering rivers with multi-grain bed material, simulating their bed deformation and sediment sorting.  The simulation results revealed the formation of particle sorting patterns in meandering rivers, with coarser and finer sediment particles accumulating near the erosional depression (scour) and depositional (point bar) regions, respectively. These findings support the coupled framework’s ability to predict sediment particle sorting and highlight the importance of representing bed material heterogeneity in numerical modeling of meandering rivers.
\end{abstract}


\end{frontmatter}

\section{Introduction}

\label{sec:intro}
\noindent Sediment transport is fundamental to morphological evolution in alluvial rivers, estuaries, and coastal environments. It shapes bedform development, influences channel stability, and contributes to the long-term adjustment of natural and engineered waterways to changing hydraulic conditions. Accurate prediction of sediment mobility supports river engineering, coastal management, infrastructure protection, benthic habitat assessment, and environmental planning that accounts for sediment dynamics, including the assessment of renewable energy projects \citep{mccarron2019hiding,sharma2019bedload,ahadi2020, selim2022}. Within this broader morphodynamic system, bedload transport plays a central role in redistributing sediment near the bed and shaping bedforms and channel morphology \citep{sharma2019bedload}. The importance of bedload transport for bed evolution computations has motivated the development of numerous theoretical and empirical formulations for predicting bedload transport.

Classic sediment transport theory was largely developed under simplified assumptions in which the bed material was either uniform or characterized by a single characteristic grain size, e.g., the median diameter $d_{50}$. Within this framework, sediment motion is commonly described using the Shields parameter, and bedload transport rates are linked to the extent to which the applied bed shear stress exceeds a critical threshold \citep{shields1936application,meyer1948formulas,einstein1950bedload, WARD201594}. The threshold-based principles provided the foundation for a range of empirical and semi-empirical bedload formulations to estimate bedload transport \cite{meyer1948formulas,einstein1950bedload,bagnold1966approach, engelund1967monograph, vanrijn1984sediment}. These models form the basis for modern sediment transport prediction and are widely used in laboratory, riverine, and coastal modeling, but their assumptions are not applicable to bed material with varying grain size \cite{SOULSBY2005673}.

In particular, the single-size representation of bed material becomes restrictive in natural environments, where bed material commonly spans a broad range of grain sizes. Representing the sediment mixture by a single representative diameter, such as $d_{50}$, cannot capture differences in mobility among individual sediment fractions. The critical bed shear stress for bed material movement and entrainment may vary among grain size classes relative to surrounding grains, their position within the sediment bed, and their interactions with neighboring particles \citep{wu2000nonuniform,mccarron2019hiding}. Therefore, a single entrainment threshold cannot fully describe the mobility of individual fractions. Consequently, formulations for uniform sediment may lose accuracy when applied to mixed-size bed material with varying local grain-size distributions \citep{wilcock2003surface}.

Cumulative laboratory studies have demonstrated the complexity of sediment mobility in multi-grain-size beds. Surface sorting and armoring regulate the exposure of different grain sizes, resulting in partial mobility among fractions of bed material under specific flow conditions \citep{parker1982gravel,wilcock2003surface}. Others have addressed the hiding and exposure effect of particles, in which coarse particles obstruct fine particles and increase the critical shear stress required for their mobilization \citep{einstein1950bedload,wu2000nonuniform,mccarron2019hiding}. For example, \citet{mccarron2019hiding} investigated the hiding-exposure effect in bimodal sand-gravel mixtures and found that the critical shear stress required to mobilize the sand fraction can increase by up to 75\%, whereas the critical shear stress required to mobilize the gravel fraction can decrease by up to 64\% compared with well-sorted sediment of similar size. 
These findings underscore the inherent complexity of sediment dynamics in natural systems with mixed-size bedload transport, showing that conventional assumptions of uniform sediment motion cannot adequately describe sediment transport over multi-grain-size bed material \cite{Duan, ZhiPeng}.

Multi-grain-size sediment transport depends on both the magnitude of the applied hydraulic forces and the interactions among grain-size fractions \cite{LEROUX, YANG2026}. This view motivates explicitly representing the mobility of individual fractions and tracking the evolving composition of the bed-material mixture. Each fraction can then be treated as a uniform bed material, for which uniform transport formulations apply. The uniform bed-material formulation therefore still provides a foundation for developing multi-grain-size sediment transport modules, allowing their integration into computational fluid dynamics (CFD) frameworks \cite{papanicolaou2008}. 

Several numerical studies have incorporated multi-grain-size sediment transport processes in their hydro-morphodynamic frameworks \cite{dargahi2004, papanicolaou2008, antze2009, tritthart2011, torok2012, ahadi2020}. For example, \citet{dargahi2004} simulated short-term sediment transport in the bifurcating Klarälven River. \citet{antze2009} predicted bed deformation by incorporating a hiding-exposure correction into multi-grain-size sediment transport by expanding the uniform model of \citet{vanrijn1984sediment}. \citet{torok2012} reproduced changes in the bed composition, including spatial patterns of coarsening and fining around a groyne structure. In a study of the Danube River in Austria \cite{tritthart2011}, the uniform Meyer-Peter and Müller (MPM) formulation \cite{meyer1948formulas} underestimated bedload transport rates at low to mean discharges. The multi-grain-size MPM formulation, which incorporates an exchange layer and is implemented in the iSed sediment transport model coupled with RSim-2D, showed reasonable agreement with field measurements. 

Despite the advances, the three-dimensional modeling of multi-grain-size sediment transport remains challenging for the computational cost and complexity of the physics. Hence, most high-fidelity CFD studies of coupled flow and morphodynamics focus on uniform bed material with a single representative particle size, such as the median grain size \cite{dallali2015, arolla2015, khosronejad2020} or a nondimensional particle-size parameter \cite{hai2013}. 

In this study, we extend our in-house Virtual Flow Solver (VFS) \cite{MHKDR_Dataset_496, [10]} to couple large-eddy simulation (LES) with a morphodynamics model that incorporates multi-grain-size sediment material and sorting processes. The framework represents the sediment mixture with multiple grain-size classes, and tracks the evolution of their local fractions and bed morphology. We note that the coupled flow and morphodynamics framework is a two-phase continuum model. In this continuum model, the bed material consists of an unlimited number of grain size classes. Given the initial composition of the bed material (i.e., the number of grain size classes and the size and fraction of each class), the multi-grain-size bed morphodynamics model predicts the evolving bed geometry and the local composition of the bed material in terms of sediment sizes and their corresponding fractions. 

To evaluate the performance of the developed multi-grain-size sediment transport model, we conducted an experimental study in a laboratory flume with multi-grain-size bed material. Using a submerged disk, the flow was locally accelerated to allow for clear water scour around the disk. The locally accelerated flow led to meaningful sediment sorting processes around the disk. Once at equilibrium, the bed geometry and sediment particle sorting were measured and utilized to validate the multi-grain-size sediment transport model. Numerical simulations using the single-grain-size sediment transport model under identical flow and geometric conditions were also performed to provide a controlled comparison of the two sediment transport formulations. After validating the developed model, we applied the multi-grain-size sediment transport framework to several large-scale meandering rivers to investigate sediment sorting within meandering rivers. In addition, we examined how multi-grain-size bed-material representation affects the magnitude and location of scour and deposition regions in large-scale river bends.

The remainder of this paper is organized as follows. \prettyref{sec:2} presents the governing equations for the hydrodynamic and morphodynamic models, describes their coupling, and introduces the multi-grain-size sediment transport framework. \prettyref{sec:3} describes the experimental setup, sediment properties, and numerical configurations used to validate the multi-grain-size sediment transport framework. \prettyref{sec:4} examines the hydrodynamic results, presents the experimental validation, compares the uniform and multi-grain-size model predictions, and discusses sediment sorting and bed morphology in meandering river geometries. Finally, \prettyref{sec:5} summarizes the study's findings.
\section{Governing equations}
\label{sec:2}
\subsection{The hydrodynamic model}
\label{sec:2.1}
\noindent The hydrodynamic model solves the spatially filtered Navier-Stokes equations for incompressible flow in non-orthogonal generalized curvilinear coordinates. Using compact index notation with the Einstein summation convention for repeated indices, the governing equations are written as follows \cite{[101], [1]}:
\begin{equation}
    J\frac{\partial U^j}{\partial\xi^j}=0
    \label{eq:1}
\end{equation}
\begin{equation}
    \frac{\partial U^i}{\partial t}=\frac{\xi_l^i}{J}\left(\frac{\partial}{\partial\xi^j}\left(U^ju_i\right)+\frac{1}{\rho}\frac{\partial}{\partial\xi^j}\left(\mu\frac{G^{jk}}{J}\frac{\partial u_i}{\partial\xi^k}\right)-\frac{1}{\rho}\frac{\partial}{\partial\xi^j}\left(\frac{\xi_i^jp}{J}\right)-\frac{1}{\rho}\frac{\partial\tau_{ij}}{\partial\xi^j}
\right)
    \label{eq:2}
\end{equation}
\noindent where $J=\left|\partial\left(\xi^1,\xi^2,\xi^3\right)/\partial\left(x_1,x_2,x_3\right)\right|$ is the Jacobian determinant of the transformation from Cartesian to generalized curvilinear coordinates. The contravariant volume flux components are defined as $U^i=(\xi_m^i/J)u_m$, where $\xi_l^i=\partial\xi^i/\partial x_l$ denotes the transformation metrics. Here, $u_i$ is the $i$th component of the filtered Cartesian velocity, $p$ denotes pressure, $\mu$ is the dynamic viscosity of water, and $\mu_{{t}}$ is the eddy viscosity. The filtered strain-rate tensor is denoted by $\overline{S}_{ij}$, and $\delta_{ij}$ is the Kronecker delta. Within the LES framework, the subgrid-scale stresses are modeled using the dynamic Smagorinsky model as follows \cite{[100], [103], [104]}:
\begin{equation}
     \tau_{ij}=-2\mu_{{t}} \overline{S}_{{ij}}+\frac{1}{3}\tau_{kk}\delta_{ij} 
     \label{eq:3}
\end{equation}
\begin{equation}
    \mu_{{t}} = C_{{s}}\Delta^2 \left |\overline{S} \right|
    \label{eq:4}
\end{equation}
\noindent where $\mu_{{t}}$ is the eddy viscosity, $\overline{S}_{ij}$ is the filtered strain-rate tensor, and $\delta_{ij}$ is the Kronecker delta. The Smagorinsky coefficient is denoted by $C_{{s}}$, and the magnitude of the filtered strain-rate tensor is defined as $\left|\overline{S}\right|=\sqrt{2\overline{S}_{ij}\overline{S}_{ij}}$. The filter width, $\Delta$, is defined as the cube root of the computational cell volume, as follows \cite{[100]}:
\begin{equation}
    \Delta = J^{-1/3}
    \label{eq:5}
\end{equation}
\noindent where $J$ represents the cell volume.

\subsection{Bed morphodynamics}
\noindent The transport of non-cohesive sediment material is modeled as bedload through rolling, sliding, and saltation, and/or as suspended load maintained by turbulence \citep{[105]}. Motion initiation and the mode of sediment transport depend on the bed shear velocity relative to the critical shear velocity and particle fall velocity. $Z_b$ denotes the elevation of the top of the bedload layer \cite{[2]}. Its temporal evolution is governed by the non-equilibrium sediment mass balance equation, known as the Exner--Polya equation \cite{[106]}:
\begin{equation}
(1-\gamma)\frac{\Delta Z_b}{\Delta t}
+\frac{1}{A_h}\sum_{e=1}^{3}
\left(\bm{q}_{BL,e}\cdot\bm{n}_e\right)
=D_b-E_b
\label{eq:6}
\end{equation}

\noindent where $\gamma$ is the sediment porosity, which is set equal to $0.4$ for the present study, and $\Delta Z_b$ is the bed elevation change over the time step $\Delta t$. The terms $D_b$ and $E_b$ denote the rates of sediment deposition from suspension onto the bed and entrainment from the bed into suspension, respectively \cite{[107]}. Because suspended sediment transport is not modeled in this study, both terms are set to zero. $A_h$ represents the horizontal projected area of the triangular bed cell, and $\bm{n}_e$ is the outward unit normal to the cell face associated with edge $e$. The edge-integrated bedload flux vector for sediment class $i$ at edge $e$ is denoted by $\bm{q}_{BL,i,e}$. The total flux vector is $\bm{q}_{BL,e}=\sum_{i=1}^{N}\bm{q}_{BL,i,e}$, where $N$ is the number of sediment grain-size classes. The bedload flux vector for bed material of class $i$ reads as \cite{[105]}:
\begin{equation}
\mathbf{\bm{q}}_{{BL},i} = {\psi}_{i} \, {\|d_{{s}}\|} \, {\|\delta_{{BL},i}\|} \, \mathbf{\bm{u}}_{{BL}}
\label{eq:7}
\end{equation}

\noindent where $d_s$ represents the edge length of triangular bed cells, $\delta_{{BL},i}$ is the bedload layer thickness corresponding to grain size class $i$, and the velocity vector within the bedload layer and parallel to the bed surface is shown with $\mathbf{\bm{u}}_{{BL}}$. 
The sediment concentration within the bedload layer for particle size class $i$, $\psi_i$, is defined as follows:
\begin{equation}
\psi_i =
\begin{cases}
0.18\,f_i C_0 \dfrac{T_i}{D_{*,i}}, & T_i>0,\\
0, & T_i\leq0.
\end{cases}
\label{eq:7_1}
\end{equation}

\noindent where $f_i$ is the local volumetric fraction of sediment class $i$ within the bed material, $C_0=0.65$ is the maximum volumetric sediment concentration
parameter adopted following the van Rijn bedload formulation \cite{vanrijn1984sediment}, and $D_{*,i}$ is the dimensionless particle diameter, defined as:

\begin{equation}
D_{*,i} = d_{i} \left[ \left( \frac{\rho_{{s}} - \rho}{\rho \nu^2}g \right)^{1/3} \right]
\label{eq:9}
\end{equation}
where $d_i$ is the representative particle diameter of sediment class $i$, $\rho_{{s}}$ is the sediment particle density, set to $2650~{kg\,m^{-3}}$, and $\nu$ is the kinematic viscosity of water. The dimensionless excess shear stress for sediment class $i$, $T_i$, reads as \cite{[105]}:
\begin{equation}
T_i = \frac{\tau_{*} - \tau_{*cr,i}^{mix}}{\tau_{*cr,i}^{mix}}
\label{eq:10}
\end{equation}

\noindent where $\tau_{*}= \rho {u_{*}}^2$ is the bed shear stress and $u_{*}$  is shear velocity. The velocity vector and the corresponding wall shear stress at the computational nodes adjacent to the solid surfaces and the water-sediment interface are determined using a wall model \cite{[101], [108]}.  $\theta_{cr0,i}$ denotes the critical Shields stress for sediment class $i$, initially evaluated for a horizontal bed as \cite{[109]}:

\begin{equation}
\theta_{cr0,i}
= \frac{0.3}{1 + 1.2D_{*,i}}
+ 0.055\left[1-\exp\left(-0.02D_{*,i}\right)\right]
\label{eq:18}
\end{equation}

\noindent The critical Shields parameter, $\theta_{cr0,50}$, is obtained as a reference value by evaluating Eq.~\eqref{eq:18} at the mixture median diameter, $d_{i}=d_{50}$. The corresponding reference value for critical bed shear stress for the flat bed, $\tau_{*cr0,50}$, is then corrected for bed-slope effects as follows \cite{[101], [110], [111]}:

\begin{equation}
\tau_{*cr,50}
=
\tau_{*cr0,50}
\left[\frac{\sin(\phi+\alpha)}{\sin(\phi)}\right],
\qquad
\tau_{*cr0,50}
=
\theta_{cr0,50}(\rho_s-\rho)gd_{50}
\label{eq:17}
\end{equation}

\noindent where $\alpha$ is the local bed slope angle at the center of each triangular bed cell, and $\phi$ is the sediment angle of repose. For the three-dimensional bed geometry, $\alpha$ is determined from the local bed slopes in the two horizontal coordinate directions and the corresponding velocity components at the top of the bedload layer \cite{[101]}. Although the angle of repose may vary slightly with particle diameter \cite{kim2018clay,zhang2019angle}, we use a constant value of $\phi=40^\circ$ for all sediment classes.

After correcting the reference critical bed shear stress for bed-slope effects, we apply an additional correction for hiding and exposure effects associated with sediment armoring. Whereas coarser particles are more exposed to the flow, while finer particles sheltered by surrounding coarser particles may gain increased resistance to entrainment. The hiding--exposure effects can be incorporated by modulating the critical bed shear stress of each grain-size class, as follows \cite{wilcock2003surface}: 

\begin{equation}
\tau_{*cr,i}^{mix}
=
\tau_{*cr,50}
\left(\frac{d_i}{d_{50}}\right)^{\beta_i},
\qquad
\beta_i
=
\frac{0.67}
{1+\exp\left(1.5-\dfrac{d_i}{d_{50}}\right)}
\label{eq:hiding_exposure}
\end{equation}

\noindent where, $\tau_{*cr,i}^{mix}$ is the critical bed shear stress of grain-size class $i$ after accounting for bed-slope and hiding--exposure effects. The reference stress, $\tau_{*cr,50}$, is the slope-corrected critical bed shear stress corresponding to the local mixture median diameter, $d_{i}=d_{50}$. The resulting corrected threshold is used to evaluate the dimensionless excess shear stress, $T_i$ (\prettyref{eq:10}).

Further, past studies often consider a constant bedload layer thickness \cite{[111],[1],[2],[5]}, however, the current study adopts a bedload layer thickness, $\delta_{BL,i}$, which varies as a function of the sediment size class and local flow conditions \cite{[105]}:
\begin{equation}
\delta_{BL,i} = 0.3\,d_{i}\,D_{*,i}^{0.7}\,T_i^{0.5}
\label{eq:bedload}
\end{equation}

The required quantities are evaluated for each grain size class at the cell centers and then interpolated to the cell faces using the second-order GAMMA differencing scheme. The resulting face values are used to calculate the class-specific bedload flux vector, $\bm{q}_{{BL},i}$. This convection scheme blends first-order upwind and second-order central differencing. Further details are provided in Ref.~\cite{[112]}.

Once the required quantities have been quantified and transferred to the cell faces, the sediment mass conservation equation (\prettyref{eq:6}). This update may produce local bed slope angles that exceed the sediment's angle of repose. A mass-conserving sand slide model is therefore applied to redistribute sediment and restore bed stability \cite{[107], [101], [112], [113], [1]}.
At each time step, we calculate local slope angles between neighboring cell centers; each interior triangular bed cell has three neighbors. Cells with slope angles exceeding the angle of repose are flagged, and excess sediment is redistributed among neighboring cells to reduce these angles while conserving mass. Because this redistribution may cause other adjacent slopes to exceed the angle of repose, the procedure is applied iteratively across all bed cells until the maximum local slope angle falls below $99\%$ of the angle of repose.

\subsection{Coupling hydrodynamics and morphodynamics}
\noindent Hydrodynamic and morphodynamic models are commonly coupled using either loose or strong coupling \cite{[108]}. In strong coupling, the boundary conditions at the sediment--water interface are updated iteratively within each time step until convergence, providing an implicit treatment of the coupling. This approach can improve numerical stability but increases computational cost. Loose coupling advances the models sequentially using interface information from the previous time step, providing an explicit treatment of the coupling without additional interface iterations. The choice of coupling strategy depends on the problem's stability and accuracy requirements and the available computational resources. In this study, loose coupling is adopted for its computational efficiency and demonstrated robustness in comparable applications \cite{[101], [107], [111], [112], [113], [1]}.

The hydrodynamic and morphodynamic models are coupled within the curvilinear immersed boundary (CURVIB) framework \cite{[100], [68], [1], [147]}. The governing equations for the flow and bed evolution are solved separately, with information exchanged at the sediment--water interface. The hydrodynamic solver (\prettyref{eq:1} to \prettyref{eq:5}) uses the updated bed elevations to define the interface geometry and the vertical bed velocities to prescribe the boundary conditions. The resulting flow velocities and bed shear stresses then feed into the morphodynamic solver (\prettyref{eq:6} to \prettyref{eq:bedload}) to calculate sediment transport and to update bed elevations.

Hydrodynamic and morphodynamic processes operate on different characteristic time scales: flow adjustments may occur over seconds to minutes, whereas bed evolution may extend over hours or days. To accommodate this disparity, a dual time-stepping approach is adopted \cite{[128]}, allowing the hydrodynamic and morphodynamic equations to advance with different computational time steps. In this study, the morphodynamic time step is set to $100$ times the hydrodynamic time step.
Global fluid mass conservation is enforced by correcting the outlet discharge at each time step to account for the inlet discharge and the rate of bed volume change, determined from successive bed elevations \cite{[110], [112]}. The outlet velocity field is first obtained through Neumann boundary conditions and then adjusted to satisfy this global mass balance. Further details are provided in Refs.~\cite{[101], [112]}.



The loose coupling approach advances the hydrodynamic and morphodynamic solvers sequentially. During coupling cycle $n+1$, the hydrodynamic equations (\prettyref{eq:1} to \prettyref{eq:5}) are solved using the bed elevations and vertical bed velocity from cycle $n$ to define the interface geometry and prescribe the bed boundary conditions. The resulting flow velocity and bed shear stress are then supplied to the morphodynamic solver (\prettyref{eq:6} to \prettyref{eq:bedload}) to update the bed elevations. The sand slide model subsequently redistributes sediment to ensure that local bed slope angles do not exceed the angle of repose. The updated bed elevations and corresponding vertical bed velocity define the interface geometry and boundary conditions for the flow solver in cycle $n+2$ \cite{[107]}.


\subsection{The multi-grain-size sediment transport framework}
\label{sec:nonuniform_framework}

\noindent The preceding subsection describes the morphodynamic formulations and algorithms we developed and successfully implemented to calculate the class-specific bedload sediment flux, $\bm{q}_{{BL},i}$, and the resulting bed elevation change. This is the primary contribution, but not the complete numerical implementation. To enable a complete replication of this approach, we also provide a detailed description of the procedures required to model a multi-grain-size sediment mixture. 

In this multi-grain-size sediment transport framework, the mobile bed is divided into $N$ grain-size classes, each characterized by a constant diameter $d_i$. The local volumetric fraction of each size class compared to other classes evolves according to the class-specific sediment flux. At each time step, using the class sediment flux, we update the fraction of each sediment class for every triangular bed cell. Algorithm~\ref {alg: nonuniform_sediment_transport} presents the update of sediment class fractions at each morphodynamic time step.

For each (mobile) triangular bed cell, the model stores the local sediment volume $\forall_{i}$ associated with each grain-size class. The total sediment volume in the cell is
\begin{equation}
\forall_{{total}}
=
\sum_{i=1}^{N}\forall{i}
\label{eq:total_class_volume}
\end{equation}
And the local volumetric fraction of class $i$ is calculated as
\begin{equation}
f_{i}
=
\frac{\forall_{i}}{\forall_{{total}}},
\label{eq:class_fraction}
\end{equation}
subject to
\begin{equation}
0\leq f_{i}\leq1,
\qquad
\sum_{i=1}^{N}f_{i}=1
\label{eq:class_fraction_constraint}
\end{equation}
These quantities are given as initial conditions for the mobile bed. The sediment concentration, bedload layer thickness, and bedload sediment flux are evaluated for each class using the formulation presented in Eqs.~\eqref{eq:7}--\eqref{eq:bedload}. The characteristic particle size of each class remains unchanged, while the local fraction of each class evolves as it has various mobility and transport rates.

Once local bedload fluxes $\bm{q}_{{BL},i}$ are calculated, the local sediment volume of each class $i$ is updated as:
\begin{equation}
\forall_{i}^{\,n+1}
=
\forall_{i}^{\,n}
-
\Delta t_m
\sum_{e=1}^{3}
\left(
\bm{q}_{{BL},i,e}
\boldsymbol{\cdot}
\bm{n}_{e}
\right)
\label{eq:class_volume_update}
\end{equation}
 where 
 $\Delta t_m$ is the morphodynamic time step, and superscripts $n$ and $n+1$ denote the current and updated time levels, respectively. With the outward-normal convention, a net outgoing flux reduces $\forall_{i}$, whereas a net incoming flux increases it. The outgoing flux should be limited by the sediment class availability at the cell. In other words, $\forall_{i}^{n+1}$ cannot be less than zero.

After all class-specific volumes are updated, $\forall_{{total}}^{\,n+1}$ is recalculated, and the updated fractions are obtained as follows:
\begin{equation}
f_{i}^{\,n+1}
=
\frac{\forall_{i}^{\,n+1}}
{\displaystyle\sum_{j=1}^{N}\forall_{j}^{\,n+1}}
\label{eq:updated_class_fraction}
\end{equation}
The fractions are constrained to the interval of $[0,1]$ and renormalized so that their sum remains equal to unity. The representative local grain diameter is subsequently calculated as
\begin{equation}
\overline{d}^{\,n+1}
=
\sum_{i=1}^{N}f_{i}^{\,n+1}d_i
\label{eq:updated_mean_diameter}
\end{equation}
We note that changes in $\overline{d}$ reflect changes in the fraction of sediment classes within a cell, rather than changes in their characteristic diameters.

Once the sediment mass-balance equation is solved and the sand-slide model corrects the bed slope, the bed elevation of each mobile bed cell is updated. At the same time, the multi-grain-size sediment transport model retains the local transport history and mixture composition of every sediment size class throughout the mobile bed domain. 

\begin{figure} [H]
\centering
  \includegraphics[width=1\textwidth]{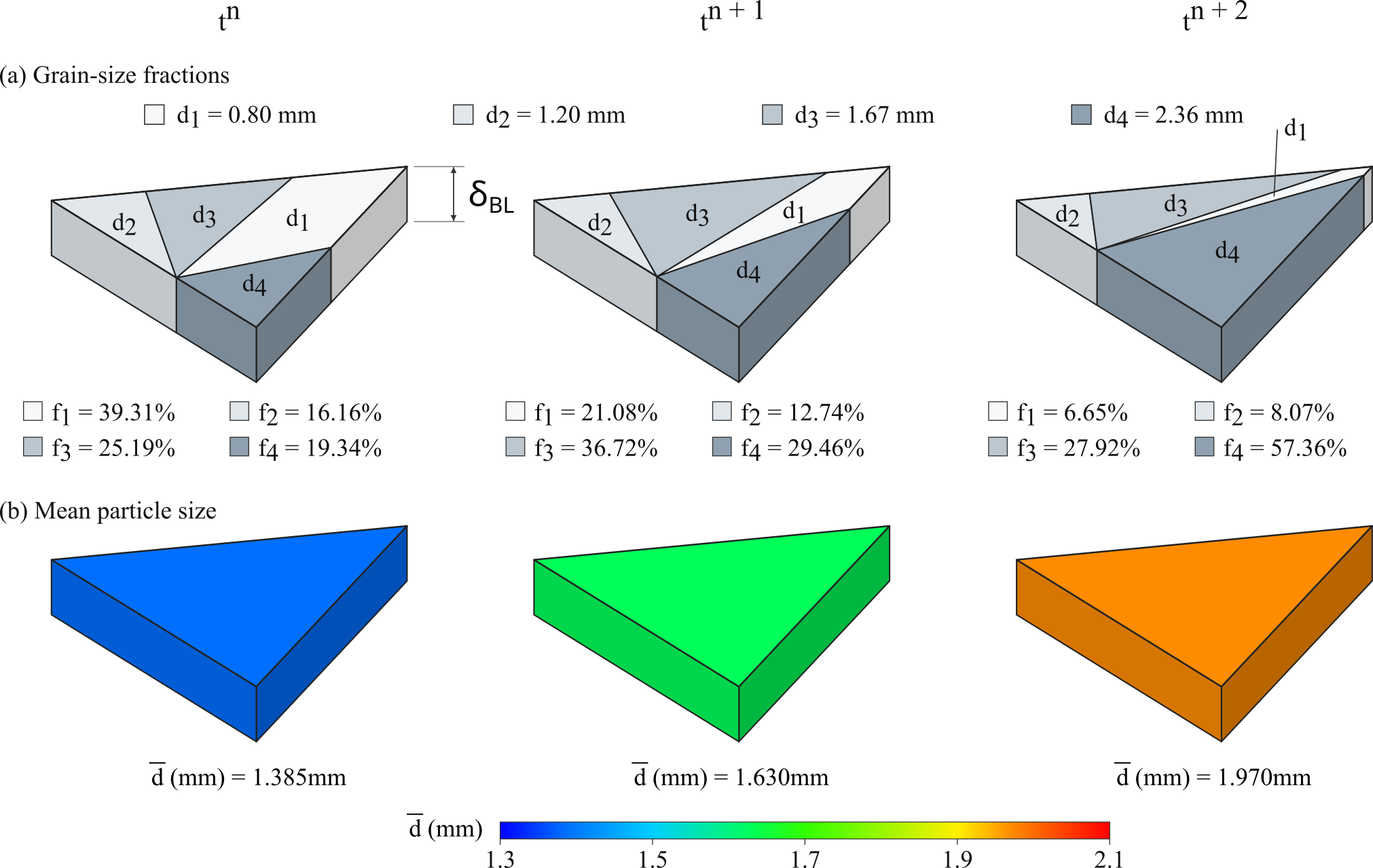}  
  \caption{Schematics of grain-size sorting within a triangular bed cell with bedload layer thickness $\delta_b$ at successive time steps $n$, $n+1$, and $n+2$. (a) shows the fractions $f_i$ of (four) grain-size classes with representative size of $d_i$; and (b) corresponds to mean particle size, $\overline{d}$, represented by the color scale in millimeters.}
  \label{fig:framework}
  \label{Fig:framework}
\end{figure}

\prettyref{fig:framework} illustrates how grain-size fractions evolve within a triangular bed cell with bedload layer thickness $\delta_{BL}$ during selective sediment transport. The sediment mixture consists of four grain-size classes, each characterized by a representative diameter $d_i$ and a local fraction $f_i$, as shown schematically in panel (a). In this illustrative example, preferential removal of finer particles increases the coarsest class's relative contribution, producing progressive local bed coarsening. At each time step, the mean particle diameter is calculated as $\overline{d}=\sum_{i=1}^{4}f_i d_i$, where $\sum_{i=1}^{4}f_i=1$. The mean diameter increases from $1.385mm$ at time step $n$ to $1.630mm$ at $n+1$ and $1.970mm$ at $n+2$. The corresponding color changes in panel (b) provide a visual representation of the evolving mean particle diameter within the cell.

\begin{algorithm}[H]
\caption{Multi-grain-size sediment transport modeling algorithm. 'N' is the total number of sediment particle size classes. }
\label{alg: nonuniform_sediment_transport}
\footnotesize
\begin{algorithmic}[1]
\Statex \textbf{Input:} Triangular grid system discretizing the mobile bed's domain; near-bed flow field; $N$ grain size classes; and initial $\forall_{i}$, $f_{i}$, and $\overline{d}$ of all bed cells.
\Statex \textbf{Output:} Class-specific sediment volumes and fractions, representative local sediment sizes, and bed elevations.
\For{each morphodynamics time step $n\rightarrow n+1$}
    \State Obtain near-bed hydrodynamics from the coupled flow-morphodynamics solver;
    \ForAll{mobile bed cells}
        \For{$i=1,\ldots,N$}
            \State Apply the hiding--exposure correction to the critical bed shear stress of the current grain-size class;
            \State Compute the class-specific $T_i$, $\psi_i$, $\delta_{{BL},i}$, and $\bm{q}_{{BL},i}$ using Eqs.~\eqref{eq:7}--\eqref{eq:bedload};

            \State Limit the outgoing flux according to the locally available amount $\forall_{i}^{n}$;
            \State Update $\forall_{i}^{n+1}=\forall_{i}^{n}-\Delta t_m\sum_{e=1}^{3}(\bm{q}_{{BL},i,e}\boldsymbol{\cdot}\bm{n}_{e})$;
            \State Enforce $\forall_{i}^{n+1}\geq0$;
        \EndFor
        \State Compute $\forall_{{total}}^{n+1}=\sum_{i=1}^{N}\forall_{i}^{n+1}$;
        \State Compute $f_{i}^{n+1}=\forall_{i}^{n+1}/\forall_{{total}}^{n+1}$ for $i=1,\ldots,N$;
        \State Enforce $0\leq f_{i}^{n+1}\leq1$ and renormalize so that $\sum_{i=1}^{N}f_{i}^{n+1}=1$;
        \State Compute $d_{{mean}}^{n+1}=\sum_{i=1}^{N}f_{i}^{n+1}d_i$;
        \State Compute the total bed-load sediment flux: $\bm{q}_{{BL}}=\sum_{i=1}^{N}\bm{q}_{{BL},i}$;
    \EndFor
    \State Use the computed $\bm{q}_{{BL}}$ to solve the sediment mass-balance equation updating the bed elevations;
    \State Correct the bed slope via sand-slide model; and
    \State Save the updated $\forall_{i}$, $f_{i}$, $d_{{mean}}$, and bed elevation.
\EndFor
\end{algorithmic}
\end{algorithm}


\section{Experimental test of multi-grain-size sediment transport }
\label{sec:3}
\begin{figure} [H]
\centering
  \includegraphics[width=1\textwidth]{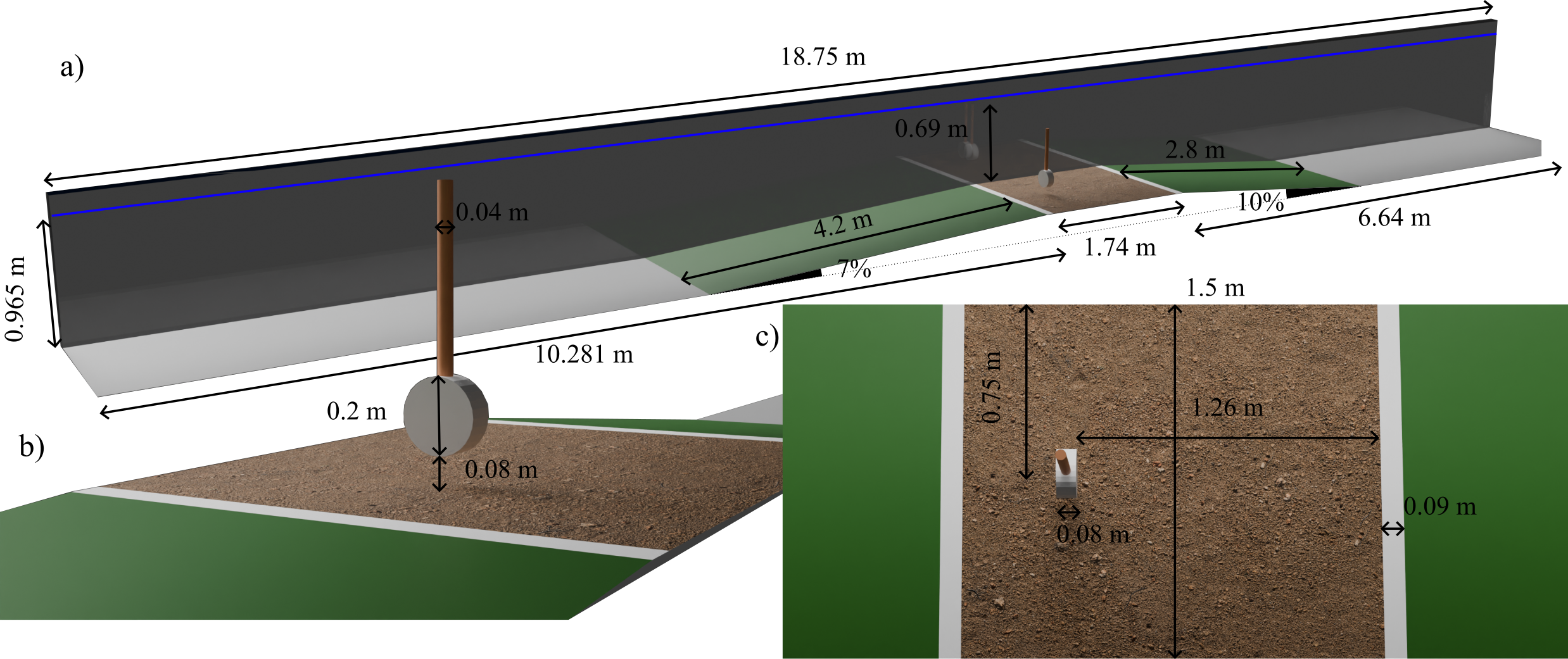}  
  \caption{Schematic of the experimental flume: (a) channel geometry, (b) disk position over the mobile bed, and (c) plan view of the mobile bed section. Flow is from left to right.}
  \label{fig:casedesc}
  \label{Fig:casedesc}
\end{figure}

\noindent The experiment was conducted in a $18.75m$ long, $1.5m$ wide, and $1.5m$ deep flume at Stony Brook University's Coastal and Hydraulic Engineering Research Laboratory \cite{[9]} (see \prettyref{Fig:casedesc}). The flume water depth was $H_{io}=0.965m$. Starting at $x=6m$ downstream of the flume inlet, the channel bed rises linearly with a $7\%$ slope, reaching a bed elevation $0.275m$ above the flume inlet elevation. The elevated region of the flume is where the mobile bed is located. At the upstream and downstream ends of the elevated mobile bed, two polyvinyl chloride (PVC) transition planks (each $0.09m$ long and $1.5m$ wide) separate the mobile bed region from the rest of the flume. The PVC planks provide a smooth transition between the rigid and mobile bed sections. The mobile bed section is $L_{{mb}}=1.74m$ long, spans the flume width, and begins at $x=10.28m$ downstream of the inlet. The water depth above the mobile bed is measured at $H=0.69m$. Downstream of the second PVC plank, the channel bed slopes down linearly at $10\%$, reaching the flume inlet bed elevation at $x=14.9m$ downstream of the channel inlet. The test was run at a bulk flow velocity of $U_{\infty}=0.5m/s$ at the flume's inlet.

\begin{figure} [H]
\centering
  \includegraphics[width=1\textwidth]{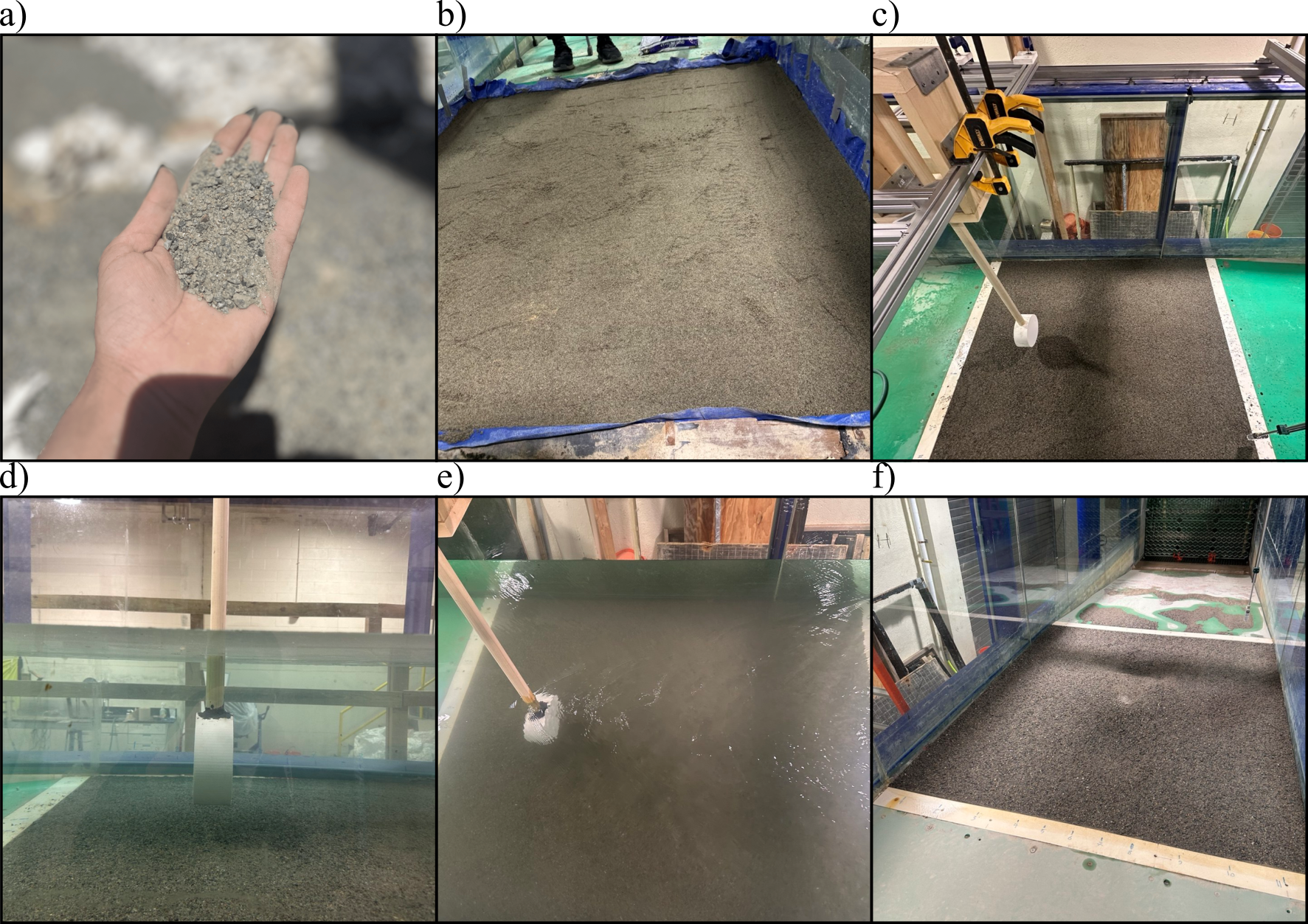}  
  \caption{The experimental setup at different testing stages: (a) sample of sediment mixture, (b) mobile bed before the test, (c) disk installation and support assembly, (d) submerged experimental setup before the test,  (e) test in progress, and (f) final bed morphology after the test. Flow in (c), (d), and (e) is from left to right.}
  \label{fig:casedesc-exp}
  \label{Fig:casedesc-exp}
\end{figure}

We utilized a solid disk to locally elevate the bed shear stress over the mobile region of the flume, inducing a clear water scour process over the mobile bed of the flume. We positioned the disk in the water column by attaching it to a shaft suspended from the top of the flume, as shown in \prettyref{Fig:casedesc}. We fabricated the disk from a high-strength plastic using three-dimensional printing. It was installed vertically, with its plane oriented normal to the primary streamwise flow. The selected material provided sufficient mechanical strength to keep the disk stable under the imposed flow conditions. The disk had a diameter of $D=0.20m$ and a thickness of $T=0.08m$, and was submerged in the water column such that its lower edge was positioned $0.08m$ above the flat mobile bed. The upstream face of the disk was located $0.40m$ downstream of the upstream edge of the mobile bed section, and the disk was centered in the spanwise direction at $0.75m$. Furthermore, we drilled a $0.045m$ diameter mounting hole in the upper portion of the disk to accommodate the $0.04m$ diameter supporting shaft. We secured the disk and shaft with waterproof adhesive. \prettyref{Fig:casedesc-exp} illustrates the experimental setup, sediment mixture, bed preparation, disk installation, submerged setup, and the bed morphology at equilibrium.


\begin{figure} [H]
\centering
  \includegraphics[width=0.8\textwidth]{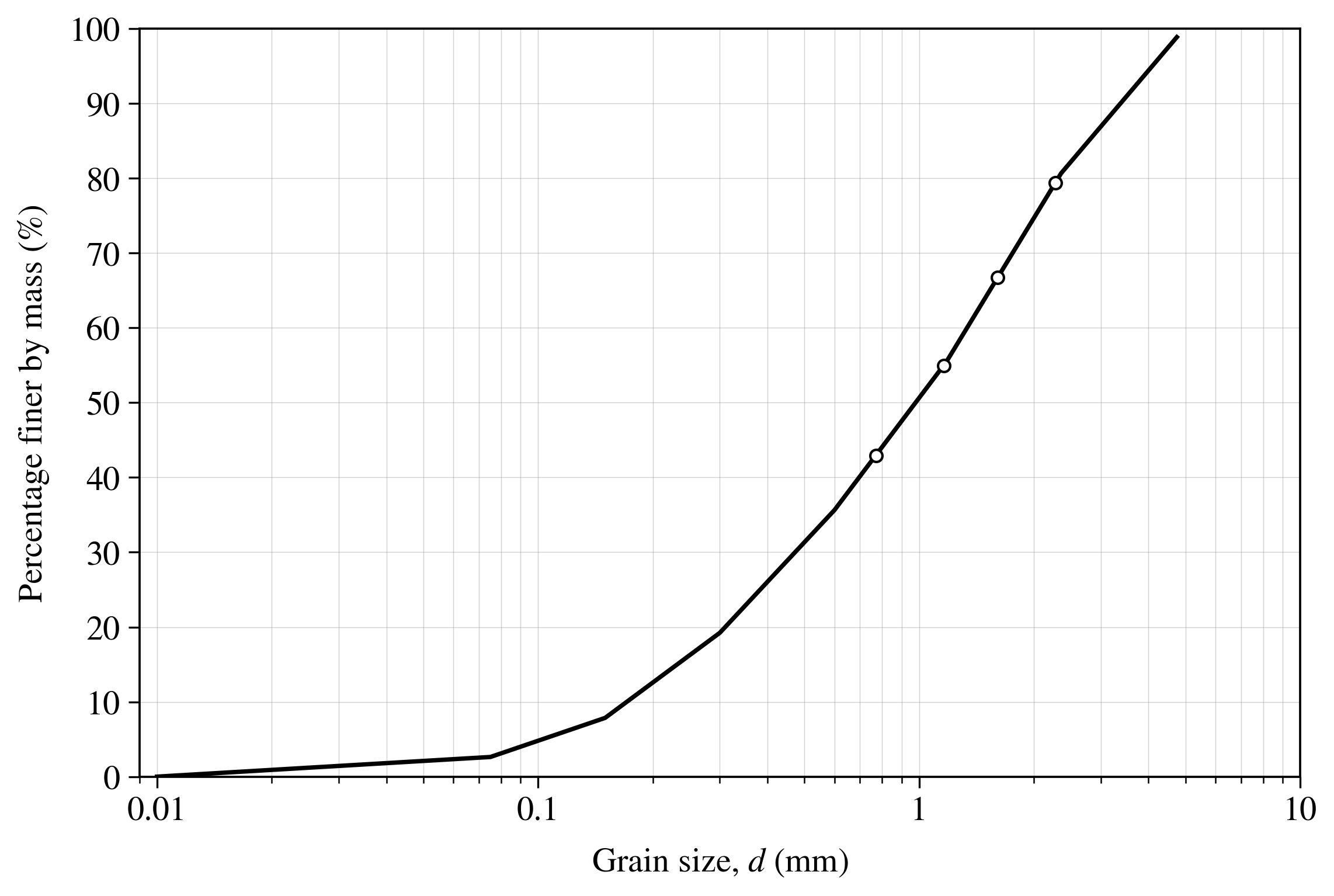}  
  \caption{Grain size distribution curve of the sediment mixture obtained from sieve analysis. The hollow circles indicate the representative diameters of the four grain-size classes used in the numerical model. Grain size is plotted on a logarithmic $x$-axis.}
  \label{fig:sieve-analysis}
  \label{Fig:sieve-analysis}
\end{figure}

In addition, the mobile bed region of the flume was filled with a sediment mixture with a density of $\rho_s=2650~{kg\,m^{-3}}$. A $30cm$ layer of the sediment mixture was employed to ensure that the deepest scour would not reach the rigid bottom of the flume. The total sediment mass placed in the mobile bed section was estimated as $M_s=1.245\times10^3~kg$. The grain size distribution of the sediment mixture is presented in \prettyref{Fig:sieve-analysis}.
Logarithmic interpolation of the measured grain size distribution yielded $d_{10}\approx0.17mm$, $d_{30}\approx0.48mm$, $d_{50}\approx0.98mm$, and $d_{60}\approx1.34mm$. These values correspond to a uniformity coefficient of $C_u=d_{60}/d_{10}\approx7.82$ and a coefficient of curvature of $C_c=d_{30}^2/(d_{10}d_{60})\approx1.01$.

According to the Unified Soil Classification System (USCS), the material is classified as well-graded sand (SW) \cite{[8]}. Table~\ref {tab:experimental_sediment} summarizes the principal properties of the experimental sediment.

\begin{table}[H]
\centering
\caption{Properties of the sediment mixture.}
\label{tab:experimental_sediment}
\begin{tabular}{ll}
\hline
\textbf{Parameter} & \textbf{Value} \\
\hline
Mobile bed depth, $D_{mb}$ (m)
& $0.30$ \\
Mobile bed volume, $V_{b} (m^3)$
& $0.783$ \\
Sediment particle density, $\rho_s (kg/m^{3})$
& $2650$ \\
Initial bed porosity, $\gamma$
& $0.40$ \\
Estimated total sediment mass, $M_s$ (kg)
& $1.245\times10^3$ \\
Angle of repose of sediment & $40^\circ$ \\
Median particle diameter, $d_{50}$ (mm)
& 0.98\\
USCS classification & Well-graded sand (SW) \\
\hline
\end{tabular}
\end{table}


Before running the turbulent flow in the flume, we leveled and scanned the sediment mixture in the mobile-bed region using a laser scanner with $1mm$ resolution.
After reaching equilibrium in the scour process, the deformed bed geometry of the mobile bed was also scanned using the laser scanner. 
The next section presents and discusses the measured bed elevation data, comparing them with the numerical simulation results. Measuring bed elevation was relatively straightforward; however, quantifying changes in sediment composition, including particle size distribution and fractional composition, within the mobile-bed region was more challenging. 

\begin{figure} [H]
\centering
  \includegraphics[width=1\textwidth]{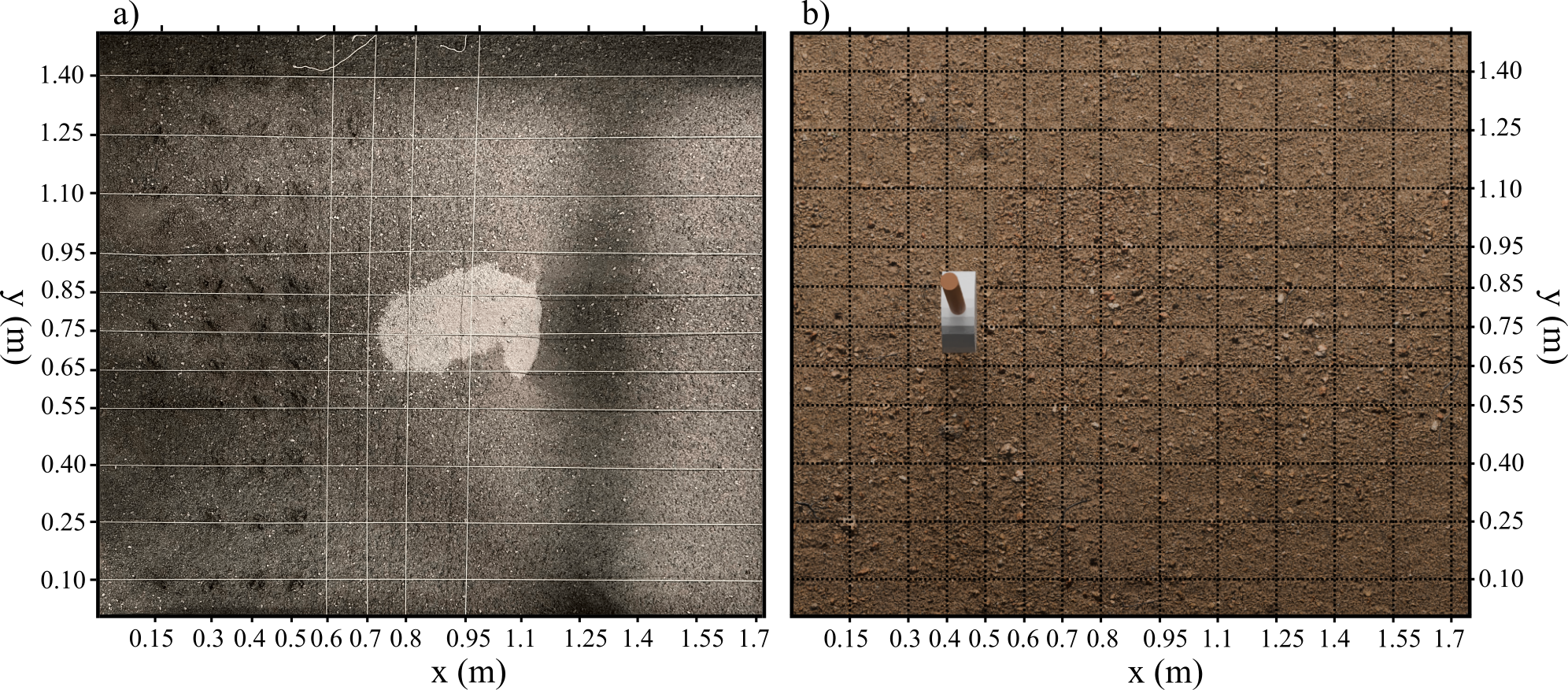}  
  \caption{Sediment sampling mesh created using threads over the mobile region of the flume (a). Sediment samples were taken at the intersection of threads, which created 143 sample nodes in (b). Flow is from left to right, and the samples were taken after the test was complete and the bed geometry reached equilibrium.}
  \label{fig:exp-post}
  \label{Fig:exp-post}
\end{figure}

To measure the changes in the sediment mixture composition and particle sorting, we first created a nonuniform grid using longitudinal and transverse threads (see \prettyref{Fig:exp-post}). The grid comprised $11$ spanwise positions and $13$ streamwise positions, yielding $143$ sampling locations. The outermost spanwise positions were located $10{cm}$ from the channel sidewalls, and the first transverse sampling row was positioned $15{cm}$ downstream of the mobile bed inlet. A finer sampling spacing of $10{cm}$ was used within a region extending $0.5D$ upstream from the disk’s upstream edge, $0.5D$ outward from each lateral edge, and $1.6D$ downstream from its downstream edge. Outside this region, we increased the sampling spacing to $15cm$. At each sampling location, nearly $10{g}$ of sediment was collected using a small hand trowel. We restricted sediment sampling to a thin layer at the bed surface to characterize flow-induced sediment sorting while minimizing the inclusion of deeper sediment that could dilute measured changes in surface grain-size composition. To access the $1.74{m}$ long sampling reach while protecting areas to be sampled, sediment collection proceeded across the channel width in successive groups of up to four transverse rows. After each group was completed, the operator stepped on the previously sampled portion of the bed and repositioned the transverse threads for the next group. This procedure continued until the final sampling row, located approximately $1.70m$ downstream of the mobile bed inlet. We first measured bed elevation with the laser scanner, then collected sediment mixture samples. We used the sediment samples for sieve analysis, and the measured sediment-sorting data (i.e., the fraction of sediment classes) are presented and discussed in comparison to the numerical simulation results in the next section.

\section{Computational details}
\noindent The computational domain includes the experimental flume, discretized by $2753\times233\times153$ grid cells respectively in the streamwise, spanwise, and vertical directions, giving a background grid spacing of $ 7mm $. We selected a time step of $1.35\times10^{-3}s$ to keep the maximum Courant--Friedrichs--Lewy (CFL) number below $1.0$ throughout the simulation. Table~\ref{tab:computational_details}. The Reynolds number based on $U_{\infty}$ and the inlet water depth was $Re_{H_{in}} \approx 4.8\times10^5$. 

\begin{table}[H]
\centering
\caption{Summary of the channel geometry, flow characteristics, and computational details.}
\label{tab:computational_details}
\begin{tabular}{ll}
\hline
\textbf{Parameter} & \textbf{Value} \\
\hline
Channel length, $L (m)$
& 18.75 \\

Channel width, $W (m)$
& 1.5 \\

Water depth at inlet, $H_{{in}}(m)$
& 0.965 \\


Water depth above mobile bed, $H (m)$
& $0.69$ \\

Length of mobile bed, $L_{{mb}}(m)$
& 1.74 \\

Length of the PVC transition $(m)$
& $0.09$ \\

Inlet velocity, $U_{\infty} (m/s)$
& $0.5$ \\

Reynolds number, $Re_{H_{in}}$
& $4.825\times10^5$ \\

Grid system dimensions
& $2753\times233\times153$ \\

Grid resolution $(mm)$
& $7$ \\

Flow solver's time step, $\Delta t (s)$
& $1.35\times10^{-3}$ \\

\hline
\end{tabular}
\end{table}

 A rigid-lid boundary condition was applied at the water surface, neglecting the fluctuation of free surfaces near the disk shaft. This seemed justified because the upper edge of the disk was $0.41m$ below the free surface, i.e., fully submerged, and did not cause any significant disturbance at the free surface.  A logarithmic velocity profile was prescribed at the inlet, and a homogeneous Neumann boundary condition corresponding to a zero normal gradient of velocity was imposed at the outlet.

As an input to the multi-grain-size sediment transport model, the sediment material, with a spectrum of sizes and fractions of each size class, had to be categorized into multiple classes. Herein, we categorized the continuous experimental grain size distribution into four representative grain size classes of $0.77$,
$1.16$, $1.61$, and $2.28{mm}$. The size intervals associated with
these representative diameters were defined by three internal class
boundaries placed midway between adjacent representative diameters on
the logarithmic grain-size axis:
\begin{equation}
    b_i = \sqrt{d_i d_{i+1}},
    \qquad i=1,2,3.
\end{equation}
This gave $b_1 \approx 0.945093mm$,
$b_2 \approx 1.366602mm$, and
$b_3 \approx 1.915933mm$.
Let $P(d)$ denote the cumulative percentage finer by mass, expressed on a
scale from $0$ to $100$. The cumulative percentage at each class boundary
was evaluated from the grain-size distribution using logarithmic
interpolation. For a boundary $b$ between two curve points $(d_L,P_L)$
and $(d_U,P_U)$,
\begin{equation}
    P(b) = P_L + (P_U-P_L)
    \frac{\ln(b/d_L)}{\ln(d_U/d_L)}.
\end{equation}
Applying this interpolation at the three class boundaries gave
$P(b_1) \approx 48.9624\%$, $P(b_2) \approx 60.8054\%$, and
$P(b_3) \approx 73.0844\%$.

The initial mass fraction of each class was then calculated from the
difference between the cumulative percentages at its upper and lower
boundaries:
\begin{equation}
\begin{aligned}
    f_{1,\mathrm{init}} &= \frac{P(b_1)}{100}, \\
    f_{2,\mathrm{init}} &= \frac{P(b_2)-P(b_1)}{100}, \\
    f_{3,\mathrm{init}} &= \frac{P(b_3)-P(b_2)}{100}, \\
    f_{4,\mathrm{init}} &= \frac{100-P(b_3)}{100}.
\end{aligned}
\end{equation}
For these class intervals, the calculated initial mass fractions were
approximately $48.96\%$, $11.84\%$, $12.28\%$, and $26.92\%$, respectively,
as summarized in Table~\ref{tab:numerical_sediment_classes}. All material
finer than the first internal boundary was included in the smallest
class, and all material coarser than the last internal boundary was
included in the largest class, ensuring that
$\sum_{i=1}^{4}f_{i,\mathrm{init}}=1$. The resulting initial mass-weighted
mean diameter, calculated using the unrounded fractions, was
$\overline{d}_{\mathrm{init}}
    = \sum_{i=1}^{4}f_{i,\mathrm{init}}d_i
    \approx 1.33{mm}$.
The hollow circles in \prettyref{Fig:sieve-analysis} mark the representative
class diameters and their vertical positions indicate cumulative percentages
finer rather than individual class fractions.

We note that this categorization can be found for every sediment sample. A greater number of sediment classes leads to a more refined representation of the sediment sample, while increasing the computational cost of the coupled flow and sediment simulations.  We argue that the four classes of particle size and their fractions considered in the computational study provide a reasonable approximation of the experimental sediment mixture. The use of four grain size classes provided a balance between representing the variability of the experimental sediment and limiting the additional computational cost associated with independently calculating the transport and local fraction of each class. The measured results of the sediment sorting are presented and discussed, in relation to the numerical modeling results, in the next section.

\begin{table}[H]
\centering
\caption{Grain size classes and initial class fractions used for the numerical simulation of the experimental test case.}
\label{tab:numerical_sediment_classes}
\begin{tabular}{cccc}
\hline
\textbf{Class} &
\textbf{Diameter (mm)} &
\textbf{Initial fraction (\%)} &
\textbf{Cumulative fraction (\%)} \\
\hline
1 & $0.77$ & $48.96$ & $48.96$ \\
2 & $1.16$ & $11.84$ & $60.8$ \\
3 & $1.61$ & $12.28$ & $73.08$ \\
4 & $2.28$ & $26.92$ & $100.00$ \\
\hline
\end{tabular}
\end{table}

\section{Results and discussions}
\label{sec:4}
\noindent We begin to discuss the results by examining the LES-computed flow field within the flume and over the rigid bed to identify the key hydrodynamic features governing sediment mobilization around the disk. Subsequently, we discuss the numerically captured morphodynamic response to the turbulent flow using both the single-grain and multi-grain-size sediment transport models. The morphodynamic analysis first focuses on validating the multi-grain-size sediment transport framework by examining the spatial distribution of particle sizes across different regions of the mobile bed surrounding the disk. Variations in sediment composition provide insight into selective transport, local sorting, and the redistribution of fine and coarse sediment fractions. Next, the morphodynamic analysis compares the computed and measured equilibrium bed geometry and bed profiles to assess the development of scour and deposition patterns. Particular attention is given to differences between the numerical results of the multi-grain-size and single-grain sediment transport solvers, highlighting how mixed-size sediment can produce distinct bed-elevation patterns through fractional mobility and size-dependent transport. These comparisons further show how particle-size variations influence both the rate and spatial structure of bed evolution.

\subsection{Instantaneous hydrodynamics}
\label{sec:rigid_bed_hydrodynamics}

\begin{figure} [H]
\centering
  \includegraphics[width=1\textwidth]{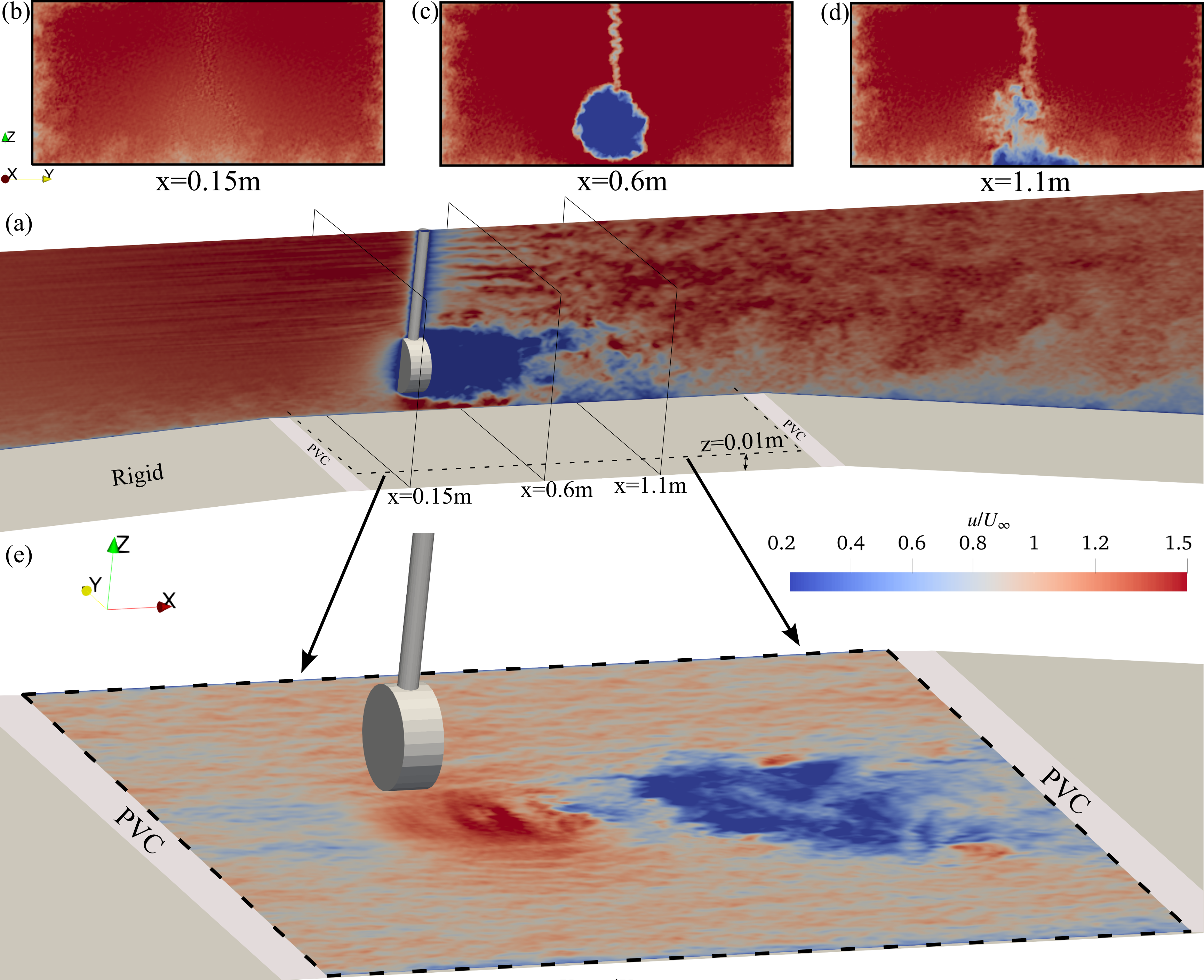}  
  \caption{Color maps of LES-computed instantaneous streamwise velocity normalized by the bulk velocity, $U_{stream}/U_{\infty}$, before allowing for bed deformation. Sagittal view over the flume's centerline is depicted in (a); transverse cross planes at $x=0.15m$, $x=0.60m$, and $x=1.10m$, respectively, are shown in (b--d); and the near-bed horizontal plane at $z=0.01m$ above the bed is presented in (e). Upstream and downstream faces of the disk are at $x=0.40m$ and $0.48m$, respectively. All streamwise locations are measured downstream from the inlet of the mobile bed section. Flow is from left to right in (a) and (e).}
  \label{fig:hydro-rigid}
  \label{Fig:hydro-rigid}
\end{figure}

\noindent We plot in \prettyref{fig:hydro-rigid}(a) contours of the instantaneous streamwise velocity normalized by the bulk velocity, $u/U_{\infty}$, over the flat mobile bed region of the flume and beyond. As shown, the flow accelerates over the upstream bed slope, generating an elevated velocity field near the entrance to the mobile bed section. This acceleration could increase the potential for sediment entrainment in the transition between the rigid and mobile bed. The submerged disk produces a pronounced velocity deficit in its wake while accelerating the flow through the $8cm$ clearance beneath its lower edge. The resulting high-momentum region lies close to the bed, indicating potential for substantial local sediment transport. The strong velocity contrast between the accelerated flow and the wake creates shear layers that are expected to promote turbulent mixing and further contribute to sediment mobilization, while the wake velocity deficit gradually decreases as the flow recovers farther downstream.

The transverse cross-sections in \prettyref{Fig:hydro-rigid}(b--d) illustrate the flow field in the wake of the disk. Considering the streamwise locations measured downstream from the inlet of the mobile bed section, at $x=0.15m$, panel (b) shows the turbulent boundary layers off of the flume sidewalls and mobile bed, immediately upstream of the disk, with low velocities near the bed and sidewalls. At $x=0.60m$, panel (c) shows a pronounced low-velocity core associated with the disk's wake, surrounded by faster flow in the shear layer and characterized by strong lateral and vertical velocity gradients. Further downstream, at $x=1.10m$, panel (d) shows a relatively weaker velocity deficit, consistent with progressive wake recovery and mixing with the surrounding flow. Nevertheless, a residual low-velocity region persists near the bed and along the channel centerline. This indicates that the near-bed flow has not yet fully recovered at this location, suggesting spatial variations in sediment-transport capacity downstream of the disk.

In contrast, \prettyref{Fig:hydro-rigid}(e) shows the near-bed instantaneous velocity field over a horizontal plane at $z=0.01m$. This panel highlights a high-velocity zone beneath and immediately downstream of the disk that is followed by a region of reduced streamwise velocity. The jet flow beneath the disk and the wake flow downstream of the disk appear to induce the high- and low-velocity regions, respectively. This velocity variability is expected to increase sediment pickup beneath the disk, reduce sediment transport capacity, and thus promote sediment deposition farther downstream. Another salient feature of the LES-captured near-bed flow dynamics includes the longitudinal turbulent streaks observed in \prettyref{Fig:hydro-rigid}(e). Such streaks have been reported to induce intermittent sweep and ejection events over mobile beds and are expected to develop at Reynolds numbers comparable to those considered in this study \cite{[13], [101], [14], [15]}. In the present simulations, these longitudinal streaks are observed on both sides of the disk, extending between the disk and the channel sidewalls.

These flow features provide a physical basis to interpret subsequent erosion and sediment sorting. For example, preferential entrainment and pickup of the finer particles beneath the disk, where near-bed shear velocity is higher, could progressively enrich the bed with coarser material. However, as the collected particles are transported downstream, where near-bed velocity is lower, reduced transport capacity could lead to deposition, and, therefore, create a downstream region enriched with finer sediment. In addition to sediment sorting, these flow patterns are expected to create localized erosion (scour) with surface coarsening beneath the disk, followed by deposition of relatively finer sediment further downstream. The hydrodynamic solver has been extensively validated elsewhere \cite{[2],[12],[68],[101], [112], [16], [128], [3], [17]}.


\subsection{Multi-grain-size morphodynamics results: grain size distribution}
\label{sec:grain_size_validation}

\noindent In this section, we compare the experimentally measured grain-size distribution of the mobile bed with the corresponding results from the multi-grain-size sediment transport model. Before presenting the results, it is important to note that the experimental measurements were obtained on a relatively coarse spatial grid, with a limited number of sampling locations (see \prettyref{sec:3} and \prettyref{Fig:exp-post} for details), whereas the computational results are available at a substantially higher spatial resolution. Thus, we generate a downsampled version of the computational results for a fair comparison. We also performed a grid sensitivity analysis to select a sufficiently fine grid resolution while keeping the computational cost manageable. The grid sensitivity analysis results are provided in Appendix~A. 

\prettyref{fig:d_exp_num} (a) to (c) show the particle size distributions of the mobile bed region of the flume at equilibrium, obtained from the experimental results and the numerical simulation results at original and downsampled resolutions, respectively. To better illustrate the temporal evolution of the computed particle size distribution, we generated an animation from instantaneous images (at original resolution) of the particle size distribution across the mobile-bed region of the channel (see Movie 1). As shown in the animation, shortly after the onset of turbulent flow over the mobile bed, the bed material begins to rearrange and gradually evolves toward a state of dynamic equilibrium.


The experimental test took nearly $ 30 min$ to reach equilibrium, while the numerical results reached equilibrium after about $ 12.6 min$ of physical time. As seen in this figure, the multi-grain-size morphodynamic model has successfully captured the primary pattern of particle size distribution, i.e., coarse and fine sediment material end up accumulating in the regions beneath and downstream of the disk, respectively. This grain-sorting pattern is consistent with the hydrodynamic interpretation in \prettyref{sec:rigid_bed_hydrodynamics}: accelerated near-bed flow beneath the disk favors preferential entrainment and downstream transport of finer particles, leaving a coarser sediment mixture beneath the disk.

After particle mobilization in the high-shear region beneath the disk, the finer particles are transported and eventually deposited farther downstream, where bed shear is more moderate. Some discrepancies between the numerical and experimental results are observed beneath the disk, where a strong jet-like flow forms. In particular, the model slightly overpredicts the particle size of sediment bed material in this region. This discrepancy could be due to ($i$) the complexity and under-resolution of the turbulent flow beneath the disk, ($ii$) uncertainties in the experimental sampling procedure, and ($iii$) our use of only four grain-size classes to represent the bed material.

\begin{figure} [H]
\centering
  \includegraphics[width=1\textwidth]{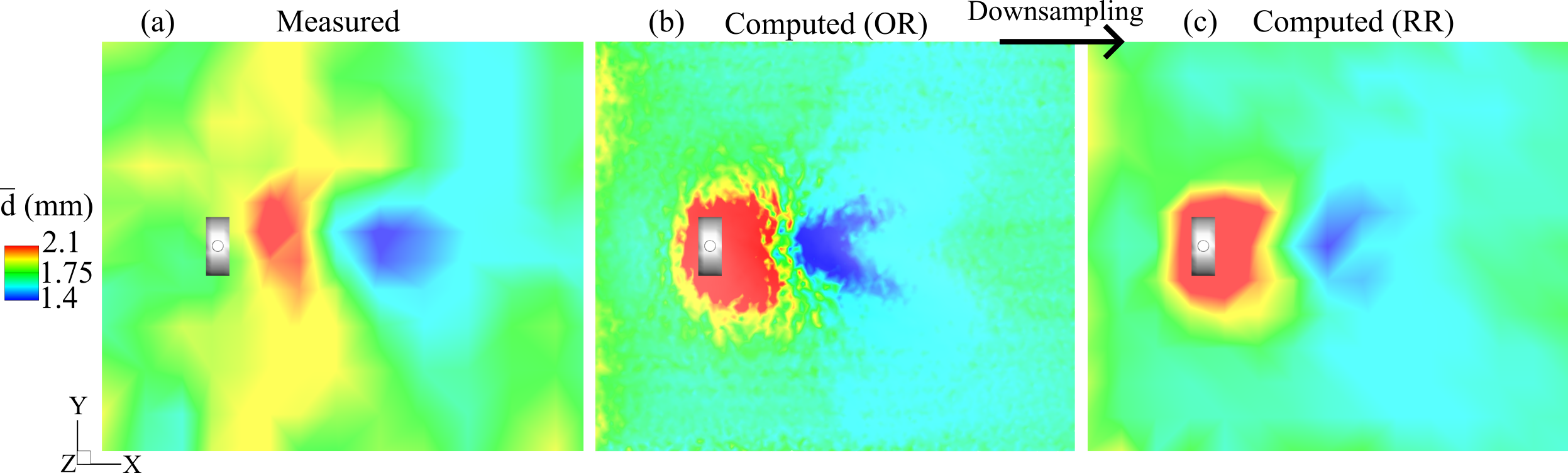}  
  \caption{Color maps of measured (a) and computed (b,c) mean sediment particle size, $\overline{d}$, in the mobile bed region of the flume. (b) presents the numerical results at the original resolution (OR), while (c) presents the downsampled version of the same numerical results at a reduced resolution (RR). The RR data are downsampled to match the experimental sampling grid of 11 by 13 in spanwise and streamwise directions, respectively. In general, red and blue represent coarse and fine sediment particles, respectively.  The flow is from left to right.}
  \label{fig:d_exp_num}
  \label{Fig:d_exp_num}
\end{figure}

\begin{figure} [H]
\centering
  \includegraphics[width=1\textwidth]{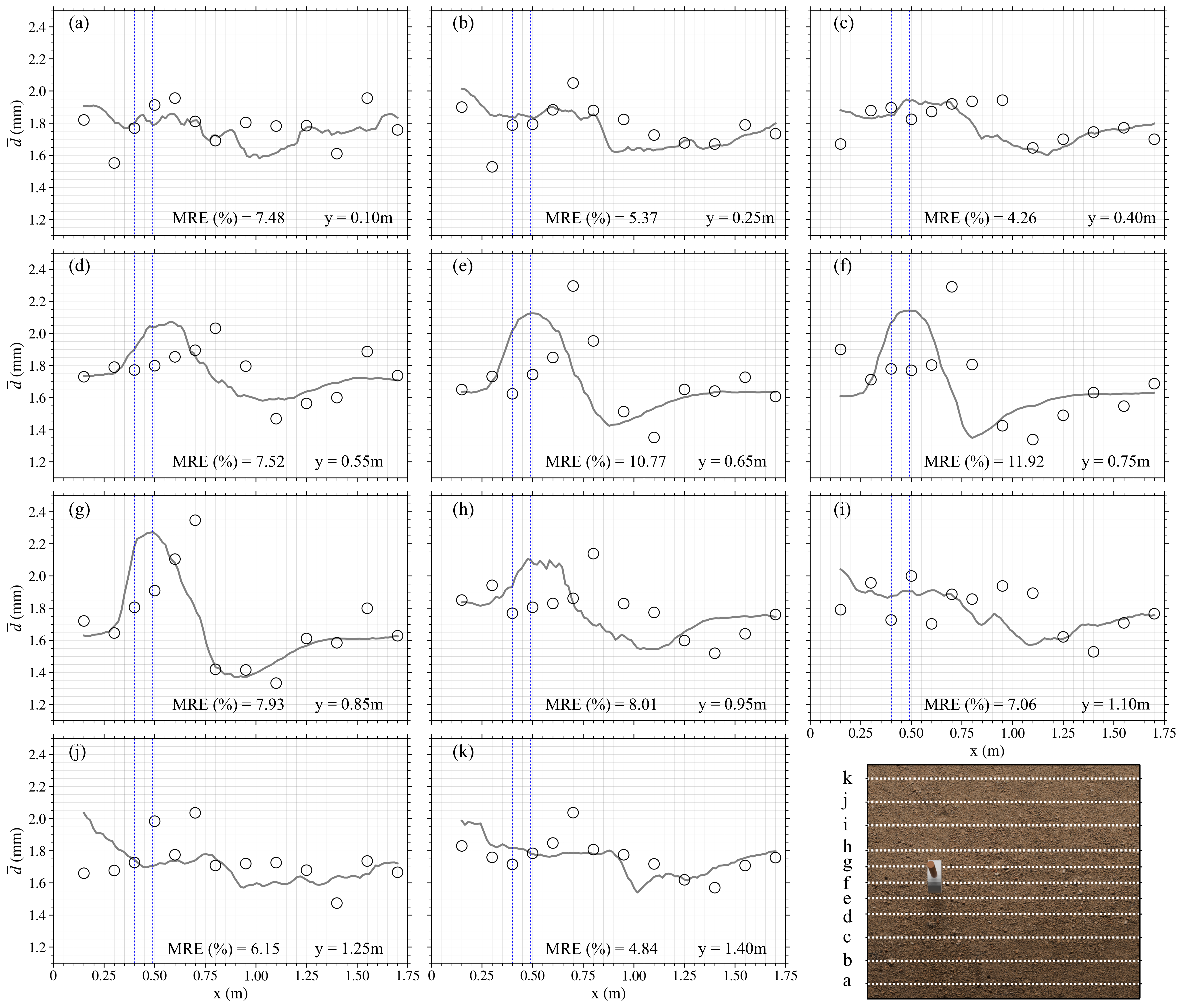} 
  \caption{Measured (circles) and computed (black line) longitudinal profiles of the mean sediment particle diameter, $\overline{d}$, at equilibrium along eleven lines shown in the bottom-right schematic of the mobile bed. The blue vertical dotted lines indicate the disk's upstream and downstream faces at $x=0.40m$ and $0.48m$, respectively. The lower-right panel shows a plan view of the mobile bed and disk, with horizontal lines indicating the streamwise profile locations corresponding to panels (a) to (k).}
  \label{fig:dmean-profile}
  \label{Fig:dmean-profile}
\end{figure}

To quantitatively assess the model's performance, \prettyref{fig:dmean-profile} shows longitudinal profiles of the mean sediment grain size from experimental measurements and numerical simulations. Overall, the comparisons demonstrate that the numerical model captures the dominant longitudinal variation in grain size across the mobile-bed region. In particular, the model reproduces the general pattern of local coarsening in the vicinity of the disk, followed by a downstream reduction in mean grain size, consistent with the redistribution of finer particles.

The most pronounced spatial gradients in grain size occur along the profiles passing through or near the disk, particularly over $y=0.55m$–$0.95m$, corresponding to \prettyref{fig:dmean-profile}(d)–(h). In this region, disk-induced flow acceleration and elevated bed shear stress promote selective sediment transport and grain-size sorting. Within the region beneath the disk, approximately $0.30 \lesssim x \lesssim 0.60$, the numerical results slightly overpredict the magnitude of the observed grain-size variation. In contrast, away from the disk, namely over $0.10 \lesssim y \lesssim 0.55m$ and $0.95 \lesssim y \lesssim 1.40m$, the numerical predictions agree more closely with the experimental measurements.

The mean relative error (MRE) quantifies the spatial variation in model accuracy for each profile in \prettyref{fig:dmean-profile}. The mean relative error ranges from $4.26\%$ to $11.92\%$, with an average of approximately $7.39\%$ across all profiles. The minimum error occurs at $y=0.40m$, whereas the maximum occurs along the disk centerline at $y=0.75m$. The elevated errors for $0.65 \lesssim y \lesssim 0.95m$ coincide with pronounced local coarsening and subsequent downstream fining, which are closely associated with disk-induced wake dynamics. This spatial pattern highlights the challenges associated with accurately resolving the coupled interactions among the flow field, bed morphodynamics, and grain-size sorting processes.

Nevertheless, the mean relative error remains below $12\%$ for all profiles, indicating reasonable overall agreement between the numerical predictions and experimental measurements throughout the mobile-bed region. Finally, the experimental dataset developed in this study provides a valuable benchmark for evaluating multi-grain-size sediment transport models. To the best of our knowledge, few, if any, experimental datasets provide spatially resolved and rigorously measured grain-size sorting data at the level of detail achieved in the present study. The dataset therefore offers a potentially unique resource for model validation and for advancing understanding of sediment-sorting processes under complex flow conditions.

\subsection{Multi-grain-size morphodynamics results: bed topography}
\label{sec:bed_elevation_validation}
\noindent We now focus on bed deformation around the disk in the mobile-bed region. We first compare the numerical predictions from the multi-grain-size sediment transport framework with the corresponding experimental measurements. To further assess the advantages of the proposed approach for considering multiple grain sizes, an additional simulation is performed using the conventional single-grain-size sediment transport model. The computed bed elevation patterns are then compared to examine the impact of multi-grain -size computations on the 
morphodynamic predictions.

\begin{figure} [H]
\centering
  \includegraphics[width=1\textwidth]{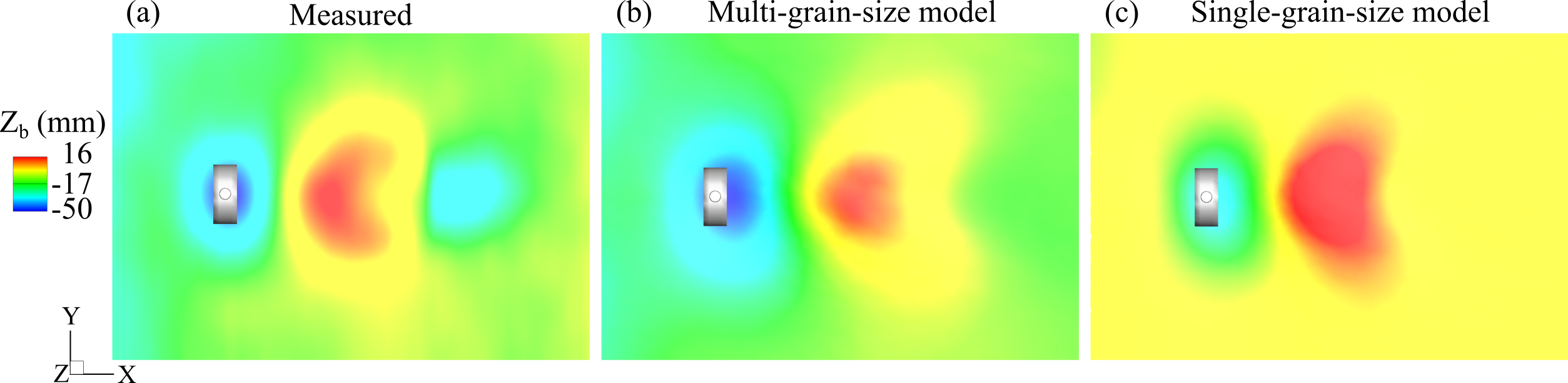}  
  \caption{Measured (a) and computed color maps of bed elevation, $Z_b$, from the top view at equilibrium. (b) and (c) correspond to the multi-grain-size and single-grain-size models, respectively. 'Red' and 'blue' mark deposition (point bar) and scour regions, respectively, while 'yellow' marks the near-flat bed (i.e., $Z_b\approx 0$) regions of the mobile bed. Flow is from left to right.}
  \label{fig:validation-zdim}
  \label{Fig:validation-zdim}
\end{figure}

\prettyref{fig:validation-zdim} shows the measured and computed bed topography of the mobile bed region after equilibrium is reached. The measurements are carried out using a sub-mm laser scanner, while the computed results are obtained via the single- and multi-grain-size bed morphodynamics models. Given the flow pattern around the disk, the following bed deformation patterns are observed in the laboratory (see \prettyref{fig:validation-zdim}(a)): ($i$) the accelerated flow field around the mounted disk leads to the main scour region beneath the disk, ($ii$) in the wake region of the disk, sediment material is deposited, creating a point bar that is similar to a single subaqueous barchan dune, at least in geometry \cite{[113]}, and ($iii$) immediately downstream of the point bar, we observed a drop in the bed elevation, owing to the point bar's wake flow dynamics, and this is the region where the fine sediment material are trapped. Furthermore, the experimentally measured maximum scour depth beneath the disk is about $5.08cm$, while the maximum deposition height of the point bar is $1.69cm$. 

Now, focusing on the bed elevation predictions of the multi-grain-size morphodynamics model, as seen in \prettyref{fig:validation-zdim}(b), the computed bed topography has a reasonable qualitative agreement with the measured topography in \prettyref{fig:validation-zdim}(a). Namely, the model has captured the scour region and the point bar's barchan dune-like geometric characteristics. The peaks of the scour depth and deposition heights are within $3.83\%$ and $11.54\%$ of the measured values, respectively. One feature that the model does not capture as accurately is the drop in bed elevation immediately downstream of the point bar. We argue that this discrepancy stems from overactivation of the sand-slide model and, perhaps, over-regulation of the bed's downhill slope immediately downstream (in the wake of) the point bar. 

More specifically, the sand-slide model used an angle of repose equal to $40^o$, which may have been too low, leading to over-regulation of the sand slopes. Namely, the angle of repose could vary with particle size and sediment composition, while the presence of cohesive fine material, such as a small clay fraction, may alter the shear strength and permit steeper stable slopes than those predicted for non-cohesive material \cite{kim2018clay,zhang2019angle}. Further, we believe that increasing the number of sediment classes would improve the accuracy of bed deformation predictions. To investigate this further, we modeled the experimental test case with a single sediment class, using $d_{50}=0.98mm$ as the representative particle size for the sediment mixture (\prettyref{tab:experimental_sediment}) as the representative particle size. 

In \prettyref{fig:validation-zdim}(c), we plot the bed topography of the mobile region at equilibrium computed by the single-grain-size morphodynamics model. Outside the disk and its wake, the flume bed elevation is nearly flat and undisturbed, $Z_b\approx 0$. This is because the sediment transport phase in the studied test case is clear-water scour, meaning the bed shear stress outside the disk region is below the critical shear stress for the single-size sediment material with $d_{50}=0.98mm$. Accordingly, the single-grain-size morphodynamics model predicts no motion initiation for particles away from the disk, and so the bed topography away from the disk is in "yellow", i.e., $Z_b\approx 0$.
This differs from the multi-grain-size model, which includes finer grain fractions; hence, finer bed material could be more readily entrained. 
Additionally, the single-grain-size model underpredicts the scour depth while overpredicting the point bar height in the wake of the disk. The error percentage of the maximum scour depth and deposition heights using this single-size model is $37.98\%$ and $48.35\%$, respectively, indicating that the multi-grain-size model is more accurate in predicting bed topography more accurately. 

Importantly, a side-by-side comparison of the mean grain-size distribution and bed elevation in \prettyref{Fig:d_exp_num} and \prettyref{Fig:validation-zdim} provides complementary evidence to assess the performance of the multi-grain-size sediment transport framework against the measured data. These figures display the measured and computed results (of the multi-grain-size model) in which thewhere relatively coarser material accumulates on the stoss side of the point bar, while relatively finer material accumulates in the wake of the point bar. This distribution is consistent with the observations of \citet{[6]}, who established that coarser sediment particles accumulate upstream of the point bar on the stoss side, while the finer sediment particles accumulate on the lee side of the point bar.

\begin{figure} [H]
\centering
  \includegraphics[width=1\textwidth]{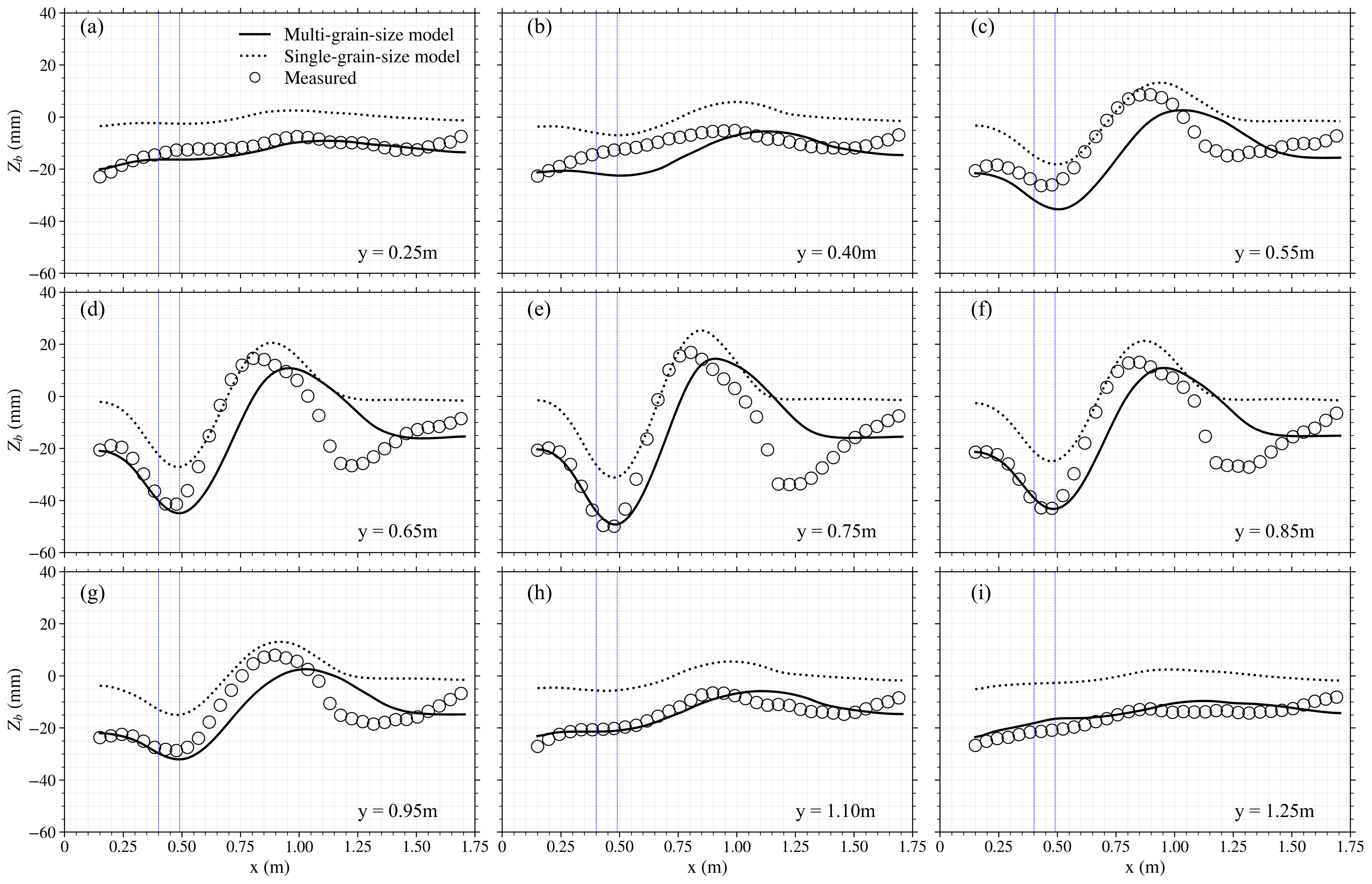}  
  \caption{Measured (circles) and computed longitudinal bed elevation profiles along the lines: (a) $y=0.25$ m, (b) $y=0.40$ m, (c) $y=0.55$ m, (d) $y=0.65$ m, (e) $y=0.75$ m, (f) $y=0.85$ m, (g) $y=0.95$ m, (h) $y=1.10$ m, and (i) $y=1.25$ m. The solid and dotted lines represent the multi-grain-size and single-grain-size sediment transport model results, respectively. The blue vertical lines indicate the disk's upstream and downstream faces, spanning $x=0.40$ m to $x=0.48$ m. The panels (a) and (b) are close to the right side wall, while the last two profiles in (h) and (i) are located near the left side wall of the flume.}
  \label{fig:z-profile}
\end{figure}

We can also compare the models' bed topography predictions with measured bed elevation profiles to produce a qualitative comparison. In \prettyref{fig:z-profile}, we compare the measured and computed longitudinal bed elevation profiles at different spanwise locations.
The physical dimensions of the laser scanner limited its measurement coverage near the flume sidewalls, so bed elevations were not recorded along the two outermost longitudinal lines at $y=0.10~m$ and $y=1.40~m$. The comparison therefore includes nine longitudinal profiles spanning $y=0.25~m$ to $y=1.25~m$.
As shown, along longitudinal profiles near the side walls (i.e., away from the scour region around the disk), the single-grain-size morphodynamics model consistently overpredicted bed elevations. In other words, the single-grain-size model did not capture much sediment transport with $d_{50}=0.98mm$. The mentioned profiles that are near the side walls include panels (a), (b), (h), and (i), where (a) and (b) are near right bank while (h) and (i) are closer to the left bank. Focusing on the bed-elevation results of the multi-grain-size model in similar regions near the side walls and away from the disk area, one can clearly observe that the model captures the bed-elevation profiles better. This is because the model considers multiple grain sizes: although the coarser particles may not move much in these regions, the finer particles are transported, leading to the observed bed deformations.

In the central disk-affected sections, particularly from approximately $y=0.55m$ to $y=0.95m$, the experimental profiles show a clear sequence of upstream and beneath-disk erosion, a depositional peak downstream of the disk, and a subsequent bed elevation drop farther downstream. The multi-grain-size sediment transport model captures this pattern significantly better than the single-grain-size model. This is specifically true about the predicted maximum scour depth beneath the disk. We argue that an important takeaway from these bed elevation predictions is that the use of the multi-grain-size morphodynamics model not only allows for quantifying the sediment size distribution and sorting, but it also significantly improves the accuracy of the bed elevation predictions. For that reason, we plan to conduct more experiments testing the validity of the notion that multi-grain-size models are more accurate than the single model in other test cases of flow-structure-sediment interactions.

Importantly, the single-grain-size sediment transport model underpredicted both the scour depth beneath the disk and the height of the point bar that developed in its wake, whereas the multi-grain-size model provided more accurate predictions of these quantities. This improvement is particularly important for simulating coupled flow and bed morphodynamics around in-stream infrastructure, including marine hydrokinetic turbine arrays and bridge piers, where inaccurate predictions of bed elevation can influence the long-term performance and stability of such infrastructure.

\section{Conclusions}
\label{sec:5}
\noindent We developed and validated a multi-grain-size sediment transport numerical framework coupled with LES to predict sediment sorting using laboratory measurements. The framework effectively models bed deformations and particle size distributions by considering multiple sediment grain-size classes. The bedload fluxes for different grain sizes were computed separately to calculate grain sorting and bed deformation. The total bed deformation was obtained by linearly summing bed changes for each grain size class. Comparisons between the modeled and measured laboratory results showed that the multi-grain-size model more accurately captures the primary patterns of local coarsening and refining of bed material, scour, and deposition. Regarding the framework's accuracy in predicting sediment sorting processes, the model's mean relative error in sediment particle size distribution was approximately $7.39\%$, with averages for individual sampling paths ranging from $4.26\%$ to $11.92\%$. The results also show that considering a range of sediment particle sizes improves the modeling accuracy. 

\begin{figure} [H]
 \centering
   \includegraphics[width=1\textwidth]{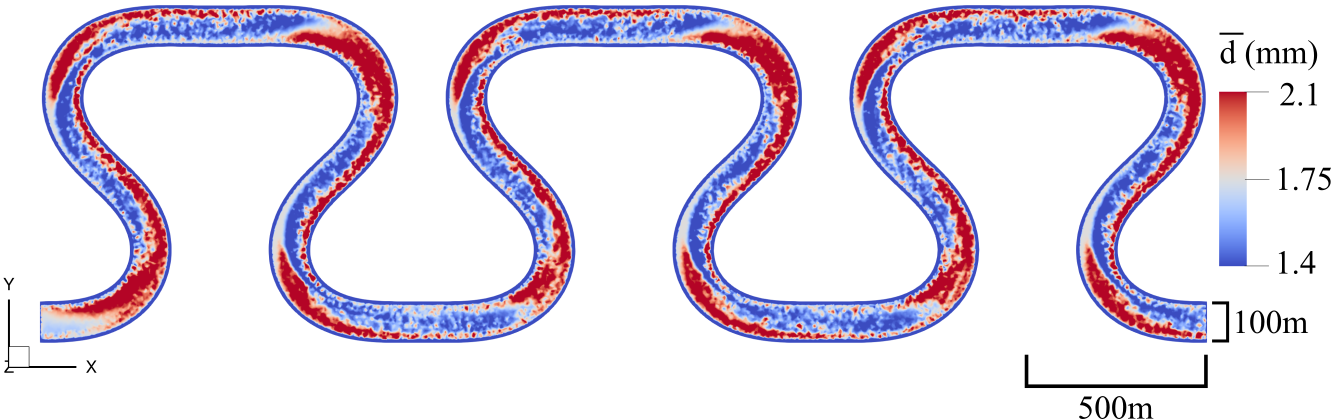}  
   \caption{Computed spatial distributions of sediment bed material, $\overline{d}$, in a large-scale compound meandering river. Red regions indicate coarser sediment, whereas blue regions indicate finer sediment. The hydraulic and geometrical characteristics of the meandering river are reported in \cite{[107]}. Flow is from left to right.}
   \label{fig:river-dmean}
   \label{Fig:river-dmean}
 \end{figure}

Importantly, the proposed multi-grain-size numerical framework can resolve spatial variations in sediment composition and provide valuable information on bed-material grain-size distribution in large-scale waterways. For example, the model can be applied to large-scale meandering rivers to characterize the spatial distribution of sediment sizes across different regions of a meander bend. Such information can improve our understanding of sediment sorting and enhance the accuracy of bed-topography predictions, which in turn inform river management and the design of infrastructure protection measures in systems where sediment sorting and local bed adjustments play important roles. As an illustrative application, we applied the coupled LES and morphodynamics model to simulate flow dynamics, bed deformation, and sediment-size distributions in a large-scale compound meandering river with a depth of 3.3 m, a width of 100 m, and a total length of 8.76 km. We specified the bed-material composition using a grain-size distribution similar to that reported in \prettyref{tab:numerical_sediment_classes}. We report details of the selected meandering-river geometry and configuration in Ref. \cite{[107]}. \prettyref{fig:river-dmean} presents the plan views of the computed bed-material grain-size distribution in the compound meandering river. The model captures spatial variability in bed-material size throughout the river and reproduces several characteristic sediment-sorting patterns. In particular, coarser and finer sediment material preferentially accumulate along the outer and inner banks of the meandering bends, respectively.

Lastly, we plan to continue research on the following aspects of the multi-grain-size sediment transport model: ($i$) performing numerical experiments to evaluate the sensitivity of the modeling results to the number of sediment classes, ($ii$) further analysis to quantify the impact of hiding and exposure parameters on the results,  ($iii$) investigating the effect of bedload layer thickness on the computations of the model, and ($iv$) examining the performance of the developed model for bridge pier scour predictions.

\FloatBarrier
\section*{Acknowledgements}
\label{sec:acknowledge}
\noindent This work was supported by grants from NSF (grant number 2233986) and the U.S. Department of Energy’s Office of Critical Minerals and Energy Innovation (CMEI) under the Hydropower and Hydrokinetic Office (H2O) Award Numbers DE-EE0009450 and DE-EE00011379. The views expressed herein do not necessarily represent the views of the U.S. Department of Energy or the United States Government. The computational resources for the simulations of this study were partially provided by the Institute for Advanced Computational Science at Stony Brook University.



\section*{Data Availability Statement}
\label{sec:7}
\noindent The software code for the VFS-$3.1$ model coupled with the multi-grain-size sediment transport framework
(\href{https://doi.org/10.5281/zenodo.23494933}{https://doi.org/10.5281/zenodo.23494933}; Ref.~\cite{[20]});
selected flow-field data
(\href{10.5281/zenodo.23516580}{10.5281/zenodo.23516580}, Ref.~\cite{[21]});
morphodynamic results at equilibrium
(\href{10.5281/zenodo.23509496}{10.5281/zenodo.23509496}, Ref.~\cite{[18]});
and experimental grain-size distribution data
(\href{10.5281/zenodo.23509326}{10.5281/zenodo.23509326}, Ref.~\cite{[19]});
associated with this study are available in the Zenodo online repository.

\appendix

\section{Grid sensitivity analysis}
\label{app:grid-sensitivity}

\noindent We performed a grid sensitivity study to select an adequate computational grid system balancing result accuracy and computational cost, using the experimental test case described in \prettyref{sec:3}. We examined three grid systems, A, B, and C, representing the coarse, intermediate, and fine grid systems, respectively, with nominal spacings of $10$, $7$, and $5mm$.  Table~\ref{tab:grid-sensitivity} summarizes the three grid systems.

\begin{table}[!htbp]
\centering
\caption{Grid systems used in the grid sensitivity assessment. The number of computational cells in the streamwise, spanwise, and vertical directions is $N_x$, $N_y$, and $N_z$, respectively. $N_B$ denotes the total for grid B.}
\label{tab:grid-sensitivity}
\small
\begin{tabular}{lrrr}
\hline
Property & Grid A & Grid B & Grid C \\
\hline
Resolution & Coarse & Intermediate & Fine \\
Nominal spacing (mm) & $10$ & $7$ & $5$ \\
Streamwise, $N_x$ & $1873$ & $2753$ & $3749$ \\
Spanwise, $N_y$ & $157$ & $233$ & $313$ \\
Vertical, $N_z$ & $101$ & $153$ & $193$ \\
Total computational cells, $N\times10^6$ & $29.70$ & $98.14$ & $226.47$ \\
Relative cell count, $N/N_B$ & $0.303$ & $1.000$ & $2.308$ \\
Compared at physical time (s) of & $30$ & $30$ & $30$ \\
\hline
\end{tabular}
\end{table}

The finest grid system (grid C) consists of approximately $226.47$ million computational nodes, compared with $98.14$ million for grid B. To limit the computational cost of coupled flow and morphodynamics modeling, we ran the simulation on grid C for $30{s}$ of physical time. We also extracted results from grids A and B at $t=30{s}$ to ensure comparisons are made at the same elapsed physical time. This assessment therefore addresses grid sensitivity during the flume's mobile-bed evolution.

\begin{figure}[H]
\centering
\includegraphics[width=\linewidth]{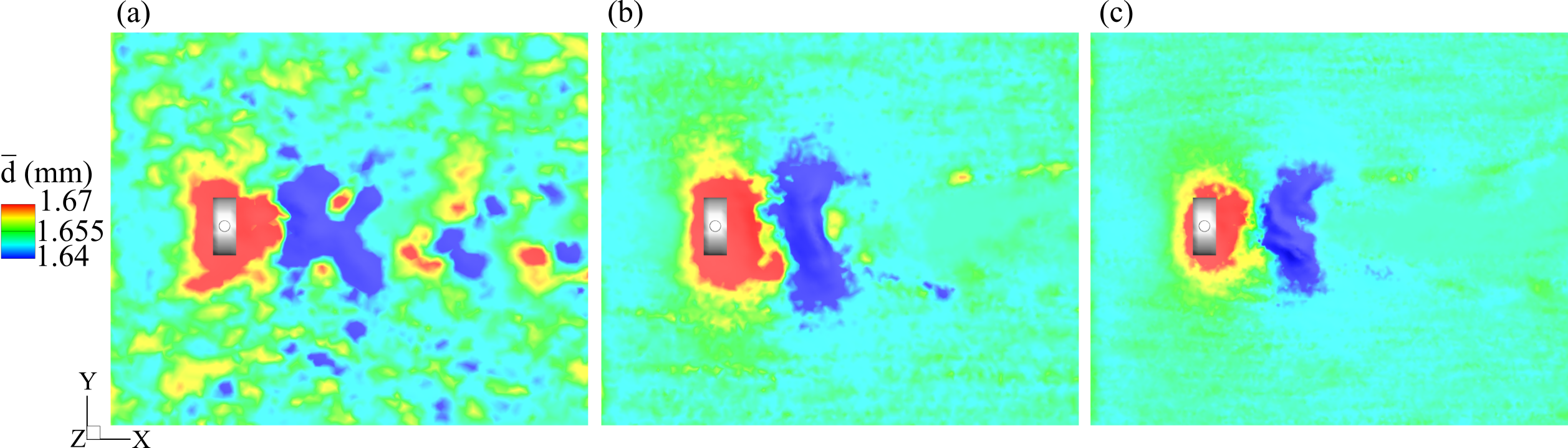}
\caption{Computed contours of spatial distributions of mean sediment diameter, $\overline{d}$ (mm), at $t=30s$ with (a) grid A ($10mm$), (b) grid B ($7mm$), and (c) grid C ($5mm$). Flow is from left to right.}
\label{fig:grid-sensitivity}
\label{Fig:grid-sensitivity}
\end{figure}

\prettyref{fig:grid-sensitivity} compares the spatial distributions of sediment grain size, $d$. Grid A exhibits fragmented sorting features and pronounced spatial variability away from the disk. Grids B and C both produce a coarser region adjacent to the disk and a finer region immediately downstream, with a similar overall spatial arrangement. Grid B predicts broader affected regions than grid C, indicating some residual grid sensitivity. Nevertheless, the two grids exhibit consistent dominant sorting patterns at the comparison time. Moreover, grid B uses $56.7\%$ fewer computational nodes than grid C, substantially reducing computational cost while preserving the principal features of the sorting response. Therefore, considering both numerical agreement and computational efficiency, we selected grid B for the study.

\bibliographystyle{elsarticle-num-names} 
\bibliography{main.bib}

\end{document}